\documentclass[aps,pra,twocolumn,showpacs,amsmath,amssymb,nofootinbib,longbibliography
]{revtex4-1}

\usepackage[pdftex]{graphicx}
\usepackage[usenames,dvipsnames]{xcolor}
\usepackage{epstopdf}
\usepackage{hyperref}
\hypersetup{
    colorlinks,
    linkcolor={black!00!black},
    citecolor={blue!100!black},
    urlcolor={blue!80!black}
}

\usepackage[normalem]{ulem}

\graphicspath{{FIGs/}}

\usepackage[utf8]{inputenc}

\usepackage{beramono}
\usepackage{listings}

\usepackage{lmodern}

\usepackage{amssymb,amsmath,amsthm}

\makeatletter
\DeclareFontEncoding{LS1}{}{}
\DeclareFontEncoding{LS2}{}{\noaccents@}
\DeclareFontSubstitution{LS1}{stix}{m}{n}
\DeclareFontSubstitution{LS2}{stix}{m}{n}
\makeatother

\DeclareMathAlphabet\mathscr{LS1}{stixscr}{m}{n}
\SetMathAlphabet\mathscr{bold}{LS1}{stixscr}{b}{n}

\makeatletter
\lst@InstallKeywords k{attributes}{attributestyle}\slshape{attributestyle}{}ld
\lst@InstallKeywords k{attributes2}{attributestyle2}\slshape{attributestyle2}{}ld
\lst@InstallKeywords k{attributes3}{attributestyle3}\slshape{attributestyle3}{}ld
\lst@InstallKeywords k{attributes4}{attributestyle4}\slshape{attributestyle4}{}ld 
\lst@InstallKeywords k{attributes5}{attributestyle5}\slshape{attributestyle5}{}ld 
\lst@InstallKeywords k{attributes6}{attributestyle6}\slshape{attributestyle6}{}ld 
\lst@InstallKeywords k{attributes7}{attributestyle7}\slshape{attributestyle7}{}ld 
\makeatother

\definecolor{macros}{rgb}{0.85,0.64,0.125}
\definecolor{types}{rgb}{0,0.55,0.55}
\definecolor{functions}{rgb}{1,0.27,0}

\definecolor{juliaterminal}{rgb}{0,0.79,0}
\definecolor{juliapkg}{rgb}{0.49,0,1}
\definecolor{juliahelp}{rgb}{0.59,0.59,0}
\definecolor{juliamodules}{rgb}{0,0.79,0.79}

\lstdefinelanguage{julia}%
  {morekeywords={abstract,break,case,catch,const,continue,do,else,elseif,%
      end,export,false,for,function,immutable,import,importall,if,in,nothing,%
      macro,module,otherwise,quote,return,switch,true,try,type,typealias,%
      using,while,mutable},%
   sensitive=true,%
   alsoother={$},%
   morecomment=[l]\#,%
   morecomment=[n]{\#=}{=\#},%
   morestring=[s]{"}{"},%
   morestring=[m]{'}{'},%
   moreattributes={@inline,@inbounds,@simd,@threads},
   attributestyle = \bfseries\color{macros}, 
moreattributes2={AbstractArray,Array,Union,Number,Integer,Tuple,NTuple,Bool,Colon,struct,Real,Float64,include,UnitRange,DataType,Function,String,StepRange},
   attributestyle2 = \bfseries\color{types}, 
moreattributes3={eigen,contract,ccontract,contractc,ccontractc,TensType,MPS,MPO,tens,Env,,intType,svd,unreshape,denstens,qarray,qr,lq,move,spinOps,makeMPO,makeMPS,expect,largeMPO,largeMPS,largeEnv,fermionOps,spinOps,tJOps,environment,makeEnv,applyMPO,largematrixproductstate,largematrixproductoperator,largeenvironment,correlation,correlationmatrix,polar,matrixproductoperator,matrixproductstate,loadMPS,loadMPO,loadLenv,loadRenv,transfermatrix,dmrg,Lupdate,Rupdate,boundaryMove,div,krylov,twosite_update,twositeOps,simpledmrg,fullH,randMPS,applyOps,applyOps,fullpsi,sub,div,add,mult,tenstype,intvecType,DMRjulia,genColType,makeId,makedens,convertTens,makeArray,convIn,findnotcons,checkType,tensorcombination,inverse_element,invmat,showTens,makepos,position_incrementer,get_ranges,pos2ind,ind2pos,maincontractor,corecontractor,prepareT,permutedims_2matrix,matrixequiv,libqr,liblq,regEnv,vecenvironment,makeoc,elnumtype,getSingleVal,joinindex,libmult,trace,checkcontract,envType,regMPS,regMPO,regEnv,tensor2disc,tensorfromdisc,compute_alpha,lanczos,compute_beta,makeEnds,densTensType,findnewm,safesvd,recursive_SVD,libsvd,libeigen,libqr,liblq,largeType,operator_in_order,moveL,moveR,movecenter,leftnormalize,rightnormalize,permutations,largeLenv,largeRenv,largeEnv,loadEnv,penalty},   
   attributestyle3 = \bfseries\color{functions},
moreattributes4={julia},
   attributestyle4 = \bfseries\color{juliaterminal}, 
moreattributes5={pkg},
   attributestyle5 = \bfseries\color{juliapkg}, 
moreattributes6={help},
   attributestyle6 = \bfseries\color{juliahelp}, 
moreattributes7={Threads,Base,LinearAlgebra,Printf,Serialization,Distributed,BLAS},
   attributestyle7 = \bfseries\color{juliamodules}, 
}[keywords,comments,strings]%

    \makeatletter
\def\@fnsymbol#1{\ensuremath{\ifcase#1\or \dagger\or \ddagger\or
   \mathsection\or \mathparagraph\or \|\or **\or \dagger\dagger
   \or \ddagger\ddagger \else\@ctrerr\fi}}
    \makeatother

\graphicspath{{FIGs/}}

\begin{document}

\title{DMRjulia (v0.8.7): Tensor recipes for entanglement renormalization computations}
\author{Thomas E.~Baker}
\email[Please send correspondence to: ]{thomasedward.baker@york.ac.uk}
\affiliation{Department of Physics, University of York, Heslington,York YO10 5DD, United Kingdom}
\affiliation{Institut quantique \& Département de physique, Université de Sherbrooke, Sherbrooke, Québec J1K 2R1 Canada}
\author{Martin P.~Thompson}
\affiliation{Institut quantique \& Département de physique, Université de Sherbrooke, Sherbrooke, Québec J1K 2R1 Canada}
\date{\today}

\begin{abstract}

Detailed notes on the functions included in the DMRjulia library are included here. This discussion of how to program functions for a tensor network library are intended to be a supplement to the other documentation dedicated to explaining the high level concepts. The chosen language used here is the high-level julia language that is intended to provide an introduction to provide a concise introduction and show transparently some best practices for the functions. This document is best used as a supplement to both the internal code notes and introductions to the subject to both inform the user about other functions available and also to clarify some design choices and future directions.

This document presently covers the implementation of the functions in the tensor network library for dense tensors. The algorithms implemented here is the density matrix renormalization group. The document will be updated periodically with new features to include the latest developments.

\end{abstract}

\maketitle


\tableofcontents



\section{Introduction}

This documentation of code is meant to be a "Numerical Recipes" style introduction to tensor network algorithms \cite{press1992numerical}.  This introduction will cover the aspects of a tensor network library from the most basic operations to the implementation of algorithms.

This code is an implementation of the concepts introduced in Refs.~\onlinecite{bakerCJP21,*baker2019m,dmrjulia1,dmrjulia2}. This introduction will focus on the details of the numerical implementation, so the interested reader is encouraged to start with the references given here if the larger physics concepts are not known.

If the user is unfamiliar with tensor networks in general, we recommend beginning with Ref.~\onlinecite{bakerCJP21,*baker2019m} which explains generally the ideas behind and uses of tensor networks. A thorough introduction to the density matrix renormalization group (DMRG), one of the most powerful original tensor network algorithms, is given in Ref.~\onlinecite{dmrjulia1}.  Implementation of this code can be found at Ref.~\onlinecite{dmrjulia}.  Additional documentation and introductory materials will be made available in the near future. Tensors with quantum number symmetries are currently available in the code, and this document will be updated with a thorough account of the functions in a subsequent update.

If the user is attempting to get some practice manipulating the basic operations in the code, it is recommended to understand the functions {\tt reshape} (Sec.~\ref{reshapefunction}), {\tt permutedims} (Sec.~\ref{permutedimsfunction}), {\tt contract} (Sec.~\ref{contractionsmodule}), and {\tt svd} (Sec.~\ref{svdfunction}) which constitute the four basic operations necessary to understand tensor network methods as outlined in Ref.~\onlinecite{bakerCJP21} and relied on heavily in Ref.~\onlinecite{dmrjulia1}. Where possible, the built-in julia functions are overloaded with functions acting on any new types that are introduced in DMRjulia, hence the code will look as close to traditional julia code as possible.

More advanced algorithms can always be broken down into these operations (with some important exceptions). So, in order to understand algorithms, the basic operations should be learned. The basic input structures from julia can be input into these functions, although some custom tensor types are implemented to reduce some extra compilation time in the code.

\subsection{Starting to use the library}

The DMRjuila library can be downloaded directly from the online repository \cite{dmrjulia}. To install the package, enter the following into the terminal to install it:

\begin{lstlisting}[numbers=none]
julia> ]
pkg> add DMRJtensor
\end{lstlisting}

To use the package, the following command will load all functions

\begin{lstlisting}[numbers=none]
julia> using DMRJtensor
\end{lstlisting}

The best option for those wishing to add different algorithms or to explore different features of a tensor network code.  To make all functions in the library available in the julia terminal or as the first line in a code using the functions here, 

\begin{lstlisting}[numbers=none]
julia> path = "<path to library>"
julia> include(path*"/DMRjulia.jl")
\end{lstlisting}

Below is an indepth discussion of the functions used in the library. For a brief description, all of the functions are implemented in DMRjulia and can be accessed in the terminal by typing (for the function {\tt contract}),
\begin{lstlisting}[numbers=none]
julia> ?
help?> DMRjulia #or other function
\end{lstlisting}

Some functions are not exported but are defined in the library. This is mainly to prevent those functions from being overwritten with new variable names in the main program and because they serve little purpose other than to aid other functions. To access these private objects after performing the above, type {\tt DMRjulia.get\_ranges} to access that function in that case. Functions that are overloaded onto a function internal to julia are immediately publicly available without any other operation as public variables.

\subsection{What paper should be cited?}

If you would like to see the original reference papers based on the code that you have written, enter the following line at the end of your code or type the following into the command line after running your code.

Depending on what type of system you are solving with DMRG, for example, it may change which papers you should cite since some papers influences how the algorithm should be used in certain situations. For example, citations for quantum chemistry and two-dimensional systems would be different.

Some general references are listed here.
\begin{itemize}
\item Two-site version of DMRG ({\tt dmrg(...,method="twosite")} \cite{white1992density,white1993density}
\item Strictly single-site (3S) DMRG ({\tt dmrg(...,method="3S")} [default]) \cite{hubig2015strictly}
\item Quantum chemistry Hamiltonians: \cite{white1999ab,chan2002highly,olivares2015ab}
\item identification of the wavefunction ansatz with a matrix product state (MPS) for DMRG \cite{ostlund1995thermodynamic,rommer1997class}
\end{itemize}

In general, it is good to familiarize oneself with the literature so that the foundational literature is always cited in addition to the specific algorithms used above \cite{bakerCJP21,dmrjulia1,schollwock2005density,schollwock2011density,white1999all,hallberg2006new,orus2014practical,bridgeman2017hand}.

This list is meant to contain the original paper where an algorithm is given. If any paper is missing here, please feel free to message the lead author with a message suggesting another paper to add to the reference list here.

The DMRjulia literature is contained in Refs.~\onlinecite{bakerCJP21,*baker2019m,dmrjulia1,dmrjulia2}. We do not anticipate that this document will be frequently cited, but if necessary, then Ref.~\onlinecite{dmrjulia0} contains the citation style. More authors are expected to be added onto this document as the library progresses.

\subsection{Debugging tips}

\subsubsection{Debugging tips}

There is a lot that can go wrong with any code, and there are a few practical tips that can help to fix most issues.

If you run into something that is not contained in these documents, can not be diagnosed, and it is not too many years since this document was published, then try contacting the author.

\begin{enumerate}
\item Focus on the top error message output by the program and work down the error messages displayed

Sometimes, although definitely not always, the error at the top of the list will be the most relevant to solving the problem.  Typically the first relevant error is contained in a file contained by the DMRjulia code itself.  Identifying which error is occurring can help to find where in the program an error is. 

\item Disable {\tt @inbounds}, {\tt @simd}, {\tt Threads.@threads} 

This can make error messages more readable and expose where segmentation faults might be.

\item Ensure that the MPS and MPO are defined with the right types.

If the computation requires complex numbers, it is best to define them at the start.

\item Design algorithms with {\tt Array}, convert to {\tt denstens}, and then start with {\tt qarray}

For ease of debugging, julia's internal array struct has many error messages for bad use.  This can be very useful in order to make the algorithm work in the first place. The {\tt denstens} is next and has less error messages.  The bulk of the errors will occur when using the quantum number system, but there will be special checks that can be added to a code to ensure its proper function.

\item Comment parts of the code until the error goes away

It typically helps to draw diagrams while the code at each step.

\item When consulting an expert in the code.

Please include the full error message and, as accurately as possible, what you were trying to do.  Other information like what machine is being used is very useful.  Even better yet, solve a problem and tell us about it to be a contributor!

\end{enumerate}

At this stage of development, the input tensors are generally the root of issues in the code. So, it is worthwhile to check twice that the inputs are programmed as needed.

It is also highly recommended to use an integrated development environment.  At the time of this document, the popular software for this is Virtual Studio Code, although Sublime, Atom, Xcode, and others are also used.

\subsection{Comment on time to run the code}

The core philosophy of the DMRjulia library is to reduce the amount of time spent implementing algorithms and learning the basics of tensor network codes. Because the implementation here is in a high-level language, there will be some noticeable speed decreases from other specific, low-level implementations of tensor networks. 

In most applications, the DMRjulia library performs as fast as other lower-level codes.  However, in other very basic examples, upwards of a 10\% slowdown has been noticed.  This is mainly for simple implementations of quantum numbers (dense tensors appear to function as fast as other implementations). Asking the user to wait a small percentrage to obtain results is prefererable, in our opinion, to requiring the user to spend more time coding an algorithm.  We will note that several highly intensive models have been run with the code (in excess of $m=10,000$ for DMRG) and results were obtained reasonably with other codes.

Explicit time comparisons will not be made here, as we think that they are not useful in the long run. The main ingredients that will help run tensor network algorithms fast is a firm knowledge of how to use those algorithms and a flexible code that can respond to what a user needs to input. The implementation here aims to help the user maintain focus on theoretical concepts instead of spending an inordinate amount of time tending to code efficiency or implementing algorithms.

\subsection{Some common implementation features in julia}

There are several implementation features in julia that deserve some discussion here to both identify why they are being used in the code and how to interpret them if implementing a tensor network library in a lower level language.

\subsubsection{Use of {\tt where} statements}

Note that there is a difference between defining the type of a variable in a function as 
\begin{lstlisting}[numbers=none]
function test(vec::Array{Number,2}
end
\end{lstlisting}
and
\begin{lstlisting}[numbers=none]
function test(vec::Array{X,2} 
	where X <: Number
end
\end{lstlisting}
The second implementation will assume all entries of the input vector {\tt vec} are the same. Meanwhile, the first implementation requires a vector input that has the element of the vector as {\tt Number} (meaning that a real number, a complex number, and an integer can be elements). This can be harder to define in the code and since there are many places where the input vector has a homogeneous type, the second implementation is used frequently.

\subsubsection{Imaginary numbers}

Imaginary numbers and complex numbers are implemented in julia.  For example, $3+4i$ would appear as {\tt 3+4im}. The library is fully compatible with imaginary numbers inside of matrices, but it is recommended to define all initial types with {\tt ComplexF64} (or similar) before starting a problem (see Sec.~\ref{MPScomplex}).

\subsubsection{Other functionality}

All other functionality not covered in this document is internal to julia.  In all cases, the library tries to look as much like julia as possible.

\subsubsection{Uniformity of input numbers and types}

Where possible, the default numbers for a function are cast as the type of input tensor to avoid any extra allocations. However, input numbers from the user are not automatically converted.  it is highly recommended to ensure that all tensor types and numbers input into functions are made to be the same type.  For example, if using complex numbers, creating the MPS, MPO, and all input values as complex will greatly reduce the amount of time needed to solve the problem.  If this is not done, then the program may not run.

\subsection{Structure of the library}

\begin{figure*}
\includegraphics[width=1.5\columnwidth]{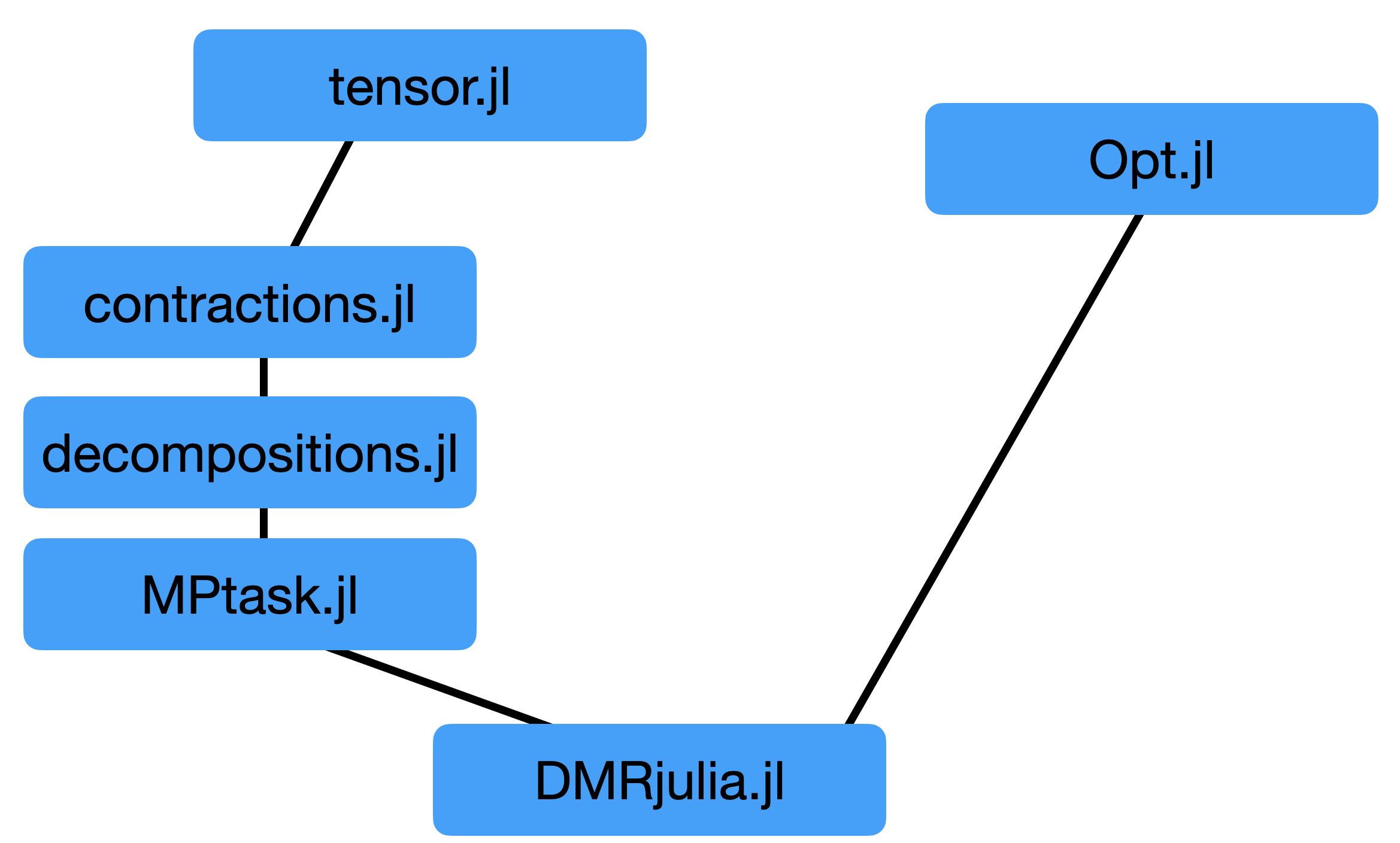}
\caption{Dependencies of the DMRjulia library.}
\label{dependencies}
\end{figure*}

An overview of how each file is dependent on the others is given in Fig.~\ref{dependencies}. Each file contains code that is related to each other.  In other implementations of the library (for example, if this code were to be implemented in a lower level language), these files would be modules containing operations for a specific purpose. In the julia language, it is convenient to simply use the {\tt include} function to load all files into one module, so the extra effort of making each file a module is skipped here.

\subsection{Some comments on the julia programming language}

Some disadvantages that will be encountered with this language in general is the tendency for julia to return lazy types on application of an operation (for example, the transpose will return an object that has instructions to compute the transpose when the program believes it is necessary to). These types can often decrease performance of the code here.  There are some other pitfalls that are remedied with the full library, but seeing how the basic {\tt Array} type works on this code will inform what is going on in the full library.  Note that all functions in the library are similar to the functions as they would be applied to regular {\tt Array} types, and the functions defined here will admit those {\tt Array} types if those are preferred over the library's defined types. For example, the apostrophe acting on a matrix gives the transpose, but the full transpose operation will not be executed immediately.

When writing code, it is recommended to write the native julia {\tt Array} type and then upgrade to the dense tensor for efficiency and backwards compatibility.

At any point in the julia terminal, once the DMRjulia is loaded, one can type ``?" and enter the function name to obtain inline comments.  The discussion here is to motivate why some functions are programmed the way that they are programmed and why other choices were made.

Whenever a function is in conflict with the pre-existing definitions in julia, it will be imported using the {\tt import} feature.

Regular programming in julia is done without declarations of what types are input into functions.  This is broken here by explicitly defining the types that are put into the julia programming langauge explicitly.  Because the normal Array type has been eschewed here, the compiler sometimes has trouble determining what output types when using custom types such as the dense tensor and quantum number tensor without these declarations. Also, including them can help with debugging.

If there is any question as to why a function was programmed in such a way, it is contained here.  The notes in the program are there for usage instructions.

There are four modifiers in lines of code that need to be discussed here. The first is the {\tt @inline} function which tells the compiler to make the function inline in the code.  This will increase the size of the executable because the program will be longer, but the time decreases. The compiler might normally do this operation for a code, but the functions where it is applied here are at the most time-sensitive elements, so it is forced here to ensure this happens.  The functions are also not very large, so will not add much to the execution time.

The second macro to keep track of is the {\tt @inbounds} flag. This means that there should be no bounds checking on the retrieval and placement of elements from an array. This can cause an overhead of 10-20\% in some cases.  The macro is only applied to the core functions.  For the most optimal performance, the global macro {\tt --check-bounds=no} which can be applied at the start of julia, will eliminate all bounds checking.

When running a simple {\tt for} loop, an optimization flag {\tt @simd} can be used to generate a small speed increase.

There is also the option to parallelize some loops with the {\tt Threads.@threads} macro.  There is a separate {\tt @distributed} macro which has a slightly different implementation.  Nested parallelization is not supported by {\tt Threads.@threads} as of the current version of julia.

At some points, the quantum number version is alluded to. All of these functions will be written for the quantum number conserving tensor. This will be added to the documentation in due time.

There are also optional arguments that can be provided to functions, and there will be an entire secret menu of inputs that can be applied to a particular function.  Whenever an input has an equals sign, the argument may be left off the function call.  In this case, the default is used.  For example, if calling {\tt dmrg}, the special option of the method can be specified, {\tt dmrg(psi,mpo,sweeps=20,method="twosite")} where {\tt sweeps} and {\tt method} are the special options here.

There is a convention in julia in-place operations to append an exclamation point (!) onto the end of a function's name.  For example, {\tt reshape} will make a copy of the input and {\tt reshape!} will change the data in-place.  Sometimes this is useful sometimes it is not.  Traditionally, the in-place function does not receive a return value.  However, it is sometimes useful for consistency to return a value anyway since returning the same value has no cost (it is returned as a pointer, not a copy).

\section{Acknowledgements}

T.E.B.~is grateful to the US-UK Fulbright Commission for financial support under the Fulbright U.S. Scholarship programme as hosted by the University of York.  This research was undertaken in part thanks to funding from the Bureau of Education and Cultural Affairs from the United States Department of State.

This library would not exist without the financial support provided to T.E.B. provided by the postdoctoral fellowship from Institut quantique and Institut Transdisciplinaire d'Information Quantique (INTRIQ). This research was undertaken thanks in part to funding from the Canada First Research Excellence Fund (CFREF).

\bibliography{TEB_papers,TEB_books,algs,refs}

\begin{widetext}

\part{Dense tensors}

\section{Main file: {\tt DMRjulia.jl} \& Overview}

The order in which the files are loaded are done in julia according to the following code.

\begin{lstlisting}[language=julia]
module DMRjulia
import LinearAlgebra #julia's interface to the BLAS linear algebra library
import Printf #julia's printing interface for outputs to the terminal
import Serialization #julia's efficient read/write of types to disk
import Distributed #julia's internal utility for parallelization

filepath = "lib/" #folder where the following files are located

include(filepath*"tensor.jl") #basic tensors
include(filepath*"Opt.jl") #operators for some models

include(filepath*"contractions.jl") #contraction functions
include(filepath*"decompositions.jl") #decompositions (eigenvalue, SVD)

include(filepath*"MPtask.jl") #making and using matrix product states/operators

filepath = "algs/" #folder containing basic files for algorithms

include(filepath*"optimizeMPS.jl") #generic optimization of an MPS
include(filepath*"Krylov.jl") #Krylov subspace expansion function
include(filepath*"DMRG.jl") #DMRG

end
using .DMRjulia
\end{lstlisting}

Each module will be discussed in depth in the following, but the first bit of code here allows for the discussion of the code strategy and to given an overview. Each set of functions that follow some theme will be labeled as a module in julia but not defined as such in the library for ease of use.  Note that by only defining the modules in julia as regular files with functions in them, all the import statements can be included here in one step, hence motivating why the formal module was not defined for each files in the following.

\section{Module: {\tt tensor.jl}}

The tensor module contains the most basic operations relating to the definition of a tensor in the code.  It is the first module defined, so it is dependent on nothing, although some functions from the {\tt LinearAlgebra} package from julia are defined.  The linear algebra functions are imported, making it so that the functions from this module must also be imported to avoid a conflict.  In all cases, the functions defined will be matched closely with the definitions in julia and this will make the code easier to use.



\subsection{Types: {\tt denstens} and general tensor types}

Some basic global types must be defined before moving onto the basic tensor type.  These types act like generic names for a wide variety of types that are defined in the following.  The export command makes the variables, types, and functions public once the library is loaded.  All other variables, types, and functions are private to the library.

\subsubsection{Type: {\tt denstens}}

The first type is a generic place-holder for the core tensor type.   To define the core tensor type, a {\tt DataType} (julia's generic type for a {\tt Float64}, {\tt ComplexF64}, etc.) corresponding to the type of number stored in the tensor.  This format allows for easier code and is named the {\tt denstens}.

\begin{lstlisting}[language=julia]
abstract type denstens end
export denstens
\end{lstlisting}

On its own, there is not much to do with this, but it will be used to define several types in the defined {\tt struct}.

\subsubsection{Type: {\tt qarray}}

This type of tensor is very different from the dense tensor ({\tt denstens}).  The quantum number conserving array will take advantage of quantum number symmetries to exploit a block symmetry in the tensor networks.  For more information, see \cite{dmrjulia2}.

\begin{lstlisting}[language=julia]
abstract type qarray end
export qarray
\end{lstlisting}

\subsubsection{Type: {\tt TensType} \& {\tt denstensType}}

To define a type that encompasses all of the tensor types possible in the library, the {\tt TensType} is defined. This can be used as a placeholder when defining a function for more than one input type.  This mainly makes coding functions easier as will be see, when they can accept more than one type, julia's principle advantage that is used here

\begin{lstlisting}[language=julia]
const TensType = Union{qarray,denstens,AbstractArray}
export TensType

const densTensType = Union{AbstractArray,denstens}
export densTensType

const tenstype = TensType
export tenstype
\end{lstlisting}

The {\tt denstensType} is useful when considering the non-quantum number types.

\subsubsection{Type: {\tt intType}}

This type is the base integer type used in the library.

\begin{lstlisting}[language=julia]
const intType = typeof(1)
export intType
\end{lstlisting}

There was no detectable difference with using another type of integer here; however, it may be useful for future users to have a common definition of the tensors here.  One could reasonably replace it with {\tt Int64} everywhere if desired. If for some reason the code is run on 32-bit architecture, then this variable can be changed to {\tt Int32} for compatibility and potential speed increases.

\subsubsection{Type: {\tt intvecType}}

Particularly for inputs for contraction, there are four possible input types that will be allowed as a generic column.  One can enter a single integer, an array of integers ({\it i.e.}, [1,2,3]), a matrix of integers with the row or column size equal to 1 ({\it i.e.} [1;2;3]), or as a {\tt Tuple}.

\begin{lstlisting}[language=julia]
const intvecType = Union{P,Array{P,1},Array{P,2},Tuple} where P <: Integer
export intvecType
\end{lstlisting}

The {\tt Tuple}s are the best as they natively carry the length of the tuple, allowing the compiler to more easily optimize the resulting code.  The {\tt Tuple} can store any type of input and can even mix inputs (for example, both integers and floats) while the traditional vector stores just one type. A {\tt Tuple}'s contents can not be changed generically once it is made.  If, for example, a {\tt Tuple} was made with a vector as one of the elements, the entries of the vector can be changed, but the size of the vector can not. This makes tuples somewhat difficult to deal with when programming, but they tend to be more efficient.  Whenever the vector does not need to change and it matches the syntax in julia's main functions, Tuples will be used, although strict attention will not be paid to it so that this code can act as a template for other languages where only vectors are used.

\subsubsection{Type: {\tt genColType}}

For operations like getting an index, there are four types of inputs that should be accommodated to match julia's inputs for the natively defined functions.

\begin{lstlisting}[language=julia]
const genColType = Union{UnitRange,Integer,Array{P,1},Colon,StepRange,
		Tuple} where P <: Integer
export genColType
\end{lstlisting}

There are four basic types of inputs, a Colon ({\tt :}), a unit range ({\it i.e.}, {\tt 1:10}), an integer, and an array of integers signifying the values to be kept on a particular index.

\subsection{Struct: {\tt tens}}

The {\tt tens} type is the core tensor type in the library.  It is a sub-type of the {\tt denstens} and carries the information of the tensor, no matter its rank.

\begin{lstlisting}[language=julia]
mutable struct tens{W <: Number} <: denstens
  size::NTuple{N,intType} where N
  T::Array{W,1}
end
export tens
\end{lstlisting}

The fields of this tensor are callable (if the {\tt tens} is {\tt A}) as {\tt A.size} and {\tt A.T} for each field respectively.  The {\tt size} field stores a tuple representing the current size of the tensor and the {\tt T} field stores the elements of a the tensor as a vector.  The number of elements in the {\tt T} field must always match the product of the elements in the {\tt size} field.

Note that there is one major difference from the {\tt Array} type in julia.  The {\tt Array} is defined with a {\tt DataType} and a rank, but since the rank is a fluid concept, it is not necessary to carry that information around in the tensor network computation.  In fact, because the rank changes so frequently, it is better to leave it out.  This will make some inefficiency with the compiler in julia, but it turns out not to be too much.  So, this tensor is defined somewhat for documentation purposes but also because it avoids the need to consider the rank in the type of a tensor.

\subsection{Constructors}

\subsubsection{Function: {\tt tens}}

There are eight ways in which were convenient to initialize a {\tt tens}.   The native constructor is to define write something like {\tt tens(W,A)} where {\tt W} is the {\tt DataType} used and {\tt A} is the input vector for the {\tt T} field. The compiler automatically determines the rest for the input function.

The first natively defined constructor is made inputting the {\tt DataType} and returning a null vector.  The tensor can be initialized as {\tt tens(), tens(Float64), tens(type=Float64)} which comprise the first three functions below.

Another option is to take a tensor defined natively in julia and convert that type into the {\tt denstens} type.  The recipe is to simply take the size from the {\tt size} function and initialize the {\tt size} field.  The tensor field can be set by reshaping the {\tt Array} into a vector.  The type of the resulting tensor can also be forced by adding it before the declaration of the {\tt Array}.

\begin{lstlisting}[language=julia]
function tens(;type::DataType=Float64)
  return tens{type}((),type[])
end

function tens(type::DataType)
  return tens{type}((),type[])
end

function tens{T}() where T <: DataType
  return tens(type=T)
end
\end{lstlisting}

When a regular julia tensor should be converted to the {\tt tens} type, the following functions will construct the tensor.

\begin{lstlisting}[language=julia]
function tens(G::DataType,P::Array{W,N}) where W <: Number where N
  sizeP = size(P)
  vecP = reshape(P,prod(sizeP))
  if G != W
    rP = convert(Array{G,1},vecP)
  else
    rP = vecP
  end
  return tens{G}(sizeP,rP)
end

function tens(P::Array{W,N}) where W <: Number where N
  return tens(W,P)
end

function tens{W}(P::Array{W,N}) where W <: Number where N
  return tens(W,P)
end

function tens{W}(P::tens{Z}) where {W <: Number, Z <: Number}
  newtens = tens(W)
  newtens.size = P.size
  newtens.T = convert(Array{W,1},P.T)
  return newtens
end
\end{lstlisting}

The conversion of the regular tensor type to a {\tt denstens} costs only a moderate amount of allocations and should not be an issue in algorithm construction. The same is true when converting back as will be seen.

If using another programming language, then converting to this type can be neglected or taken from the native type in that language.

\subsubsection{Function: {\tt rand}}

It is sometimes useful to generate a random tensor.  This version simply uses the julia version and converts to the dense tensor type.

\begin{lstlisting}[language=julia]
import Base.rand
function rand(rho::tens{W}) where W <: Number
  return tens{W}(rand(W, size(rho)))
end
\end{lstlisting}

\subsubsection{Function: {\tt zero}}

Similar to the julia function {\tt zero}, a tensor with all zero entries can be created from a pre-existing dense tensor.

\begin{lstlisting}[language=julia]
import Base.zero
function zero(M::tens{W}) where W <: Number
  return tens{W}(zeros(W,M.size))
end
\end{lstlisting}

\subsubsection{Function: {\tt makeId}}

This function generates an identity tensor.  This is most commonly used when contracting two indices on a tensor ({\it i.e.}, tracing out an index).

\begin{lstlisting}[language=julia]
function makeId(W::DataType,ldim::Integer;addone::Bool=false,
		addRightDim::Bool=false)
  oneval = W(1)
  if addone
    if addRightDim
      newsize = (ldim,ldim,1)
    else
      newsize = (1,ldim,ldim,1)
    end
  else
    newsize = (ldim,ldim)
  end
  Id = zeros(W,ldim^2)
  @simd for i = 1:ldim
    @inbounds Id[i + ldim*(i-1)] = oneval
  end
  return tens{W}(newsize,Id)
end

function makeId(A::tens{W},iA::R) where R <: intvecType where W <: Number
  Id = makeId(W,size(A,iA[1][1]),addone=true,addRightDim=true)
  for g = 2:length(iA)
    addId = makeId(W,size(A,iA[g][1]),addone=true,addRightDim=false)
    Id = contract(Id,ndims(Id),addId,1)
  end
  newsize = ntuple(a->size(Id,a),ndims(Id)-1)
  return reshape!(Id,newsize...)
end

function makeId(A::Array{W,N},iA::R) where N where {R <: intvecType, W <: Number}
  densA = tens(A)
  Id = makeId(densA,iA)
  return makedens(Id)
end
export makeId
\end{lstlisting}

\subsubsection{Function: {\tt convertTensor}}

This function converts the element type of the dense tensor into another type (from the type of {\tt M} to a type {\tt G}).

\begin{lstlisting}[language=julia]
function convertTens(W::DataType, M::denstens)
  return tens{W}(M.size,convert(Array{W,1}, M.T))
end
export convertTens
\end{lstlisting}

This is slightly contrasted with julia's conversion function ({\tt convert}), but it is easier to define and use.  It is also a rarely used function, so it is recommended that if the data type must be changed, that this function should be read about and searched for.  Often, it is much better to define all types at the beginning of a computation and continue from there instead of converting between tensor types.

\subsubsection{Function: {\tt makedens}}

The function {\tt makedens} converts a {\tt denstens} to the {\tt denstens} format (trivial operation).

\begin{lstlisting}[language=julia]
function makedens(M::denstens)
  return M
end

function makedens(M::AbstractArray)
  return tens(M)
end
export makedens
\end{lstlisting}

While function does nothing for now, it is used later for quantum number tensors and having the unified interface can be useful in several situations.

\subsubsection{Function: {\tt makeArray}}

The function {\tt makedens} converts a {\tt denstens} to the julia {\tt Array} format.

\begin{lstlisting}[language=julia]
function makeArray(M::denstens)
  return reshape(M.T, size(M))
end

function makeArray(M::AbstractArray)
  return M
end
export makeArray
\end{lstlisting}

\subsection{Helper functions}

\subsubsection{Function: {\tt convIn}}

Converts any input of type {\tt intvecType} to a Tuple.  This is most useful for contractions, but it is used in many places in the library, so it is defined in {\tt tensor}.

\begin{lstlisting}[language=julia]
function convIn(iA::Union{Array{P,1},Array{P,2}}) where P <: Integer
  return ntuple(i->iA[i],length(iA))
end

function convIn(iA::Integer)
  return (iA,)
end

function convIn(iA::NTuple{N,intType}) where N
  return iA
end
export convIn
\end{lstlisting}

The tuples here carry the information about how many elements there are in the code. The conversion of the array types to the tuple will force the code to determine how long the tuple will be. So, the use of only the tuples (and integers) as input to this function will ultimately be slightly faster, but the overall performance increase with this change is very minor.

\subsubsection{Function: {\tt findnotcons}}

Generates the complement set of an input.  This is used for contractions, but it has uses in other functions that mandate its definition be here.  For example, the complement of {\tt (1,2,6)} for a total rank of 10 would be {\tt (3,4,5,7,8,9,10)}.

\begin{lstlisting}[language=julia]
function findnotcons(nA::Integer,iA::NTuple{N,intType}) where N
  notcon = ()
  for i = 1:nA
    k = 0
    notmatchinginds = true
    while k < length(iA) && notmatchinginds
      k += 1
      notmatchinginds = iA[k] != i
    end
    if notmatchinginds
      notcon = (notcon...,i)
    end
  end
  return notcon
end
export findnotcons
\end{lstlisting}

\subsubsection{Function: {\tt checkType}}

If one of the two inputs is not matched ({\it i.e.}, an {\tt Array} and a {\tt denstens}), then the {\tt Array} is converted to a {\tt denstens}.

\begin{lstlisting}[language=julia]
function checkType(A::R,B::S) where {R <: AbstractArray, S <: denstens}
  return tens(A),B
end

function checkType(A::S,B::R) where {R <: AbstractArray, S <: denstens}
  return A,tens(B)
end

function checkType(A::AbstractArray,B::AbstractArray)
  return A,B
end

function checkType(A::qarray,B::qarray)
  return A,B
end

function checkType(A::denstens,B::denstens)
  return A,B
end
export checkType
\end{lstlisting}

Note that the last function call is between two tensor types which share a type. The cost to evaluating this function is effectively removed if two identical tensor types are input into the function, and this will be the function used by the julia language when two identical tensor types are present.

\subsubsection{Function: {\tt ind2pos!}}

This function converts a set of positions in indices to a position in a tensor. The essential operations that accomplish this are to first envision the tensor as a matrix of size of the first dimension of the first index by the rest of the indices reshaped together. Then, the same idea to find the row and column of an element in a matrix may be used to find the positions in the tensor. 

First, the index is converted to the 0-indexed version (julia uses 1-indexed numbers, so the value is subtracted by 1). Then, the value is divided by the dimension of the first index and rounded down, $\lfloor x/S_j\rfloor$ for index $x$ and size of an index $j$, $S_j$.  The column (first index's) position is now known. To find the row value, the modulus of the index is taken and a one is added since the value is 1-indexed here, $x\mod S_j$.

\begin{lstlisting}[language=julia]
@inline function ind2pos!(currpos::Array{Array{X,1},1},k::X,x::Array{X,1},
				index::X,S::Array{X,1}) where X <: Integer
  @inbounds currpos[k][1] = x[index]-1
  @simd for j = 1:size(S,1)-1
    @inbounds currpos[k][j+1] = fld(currpos[k][j],S[j])
    @inbounds currpos[k][j] = (currpos[k][j]) % S[j] + 1
  end
  @inbounds currpos[k][size(S,1)] = currpos[k][size(S,1)] % S[size(S,1)] + 1
  nothing
end
\end{lstlisting}

The function is often used to identify index values for quantum number tensors.

\subsubsection{Function: {\tt pos2ind!}}

It is frequently the case that an array containing the position in a tensor will need to be converted to a number and viceversa.  This version of the function, {\tt pos2ind!}, performs this operation in place with pre-existing vectors.

The first implementation of this function returns a single scalar value.

\begin{lstlisting}[language=julia]
@inline function pos2ind(currpos::NTuple{N,P},S::NTuple{N,P}) where {N, P <: Integer}
  x = 0
  @simd for i = length(S):-1:2
    @inbounds x += currpos[i]-1
    @inbounds x *= S[i-1]
  end
  @inbounds x += currpos[1]
  return x
end
export pos2ind

@inline function pos2ind!(x::Array{X,1},j::Integer,
			currpos::Union{Array{X,1},NTuple{N,intType}},
			S::NTuple{N,intType}) where {N, X <: Integer}
  @inbounds x[j] = currpos[end]
  @simd for i = length(S)-1:-1:1
    @inbounds x[j] -= 1
    @inbounds x[j] *= S[i]
    @inbounds x[j] += currpos[i]
  end
  nothing
end
export pos2ind!
\end{lstlisting}

The second function admits a predefined vector to eliminate any extra allocations from the first implementation.  It can be noted that the definition of a single scalar value is very cheap in terms of both computational time and garbage collection. However, care is taken to allow for pre-defined vectors to be input into functions to avoid defining vectors inside of functions that may be called many times.

Use of the {\tt @inbounds} and {\tt simd} flags is to reduce runtime but often has only a modest effect.

The function is most often useful for searching for an element of a tensor in the {\tt denstens} format. By converting the position to an index, it can make it easier to search for an element in a tensor of the {\tt denstens} format.

\subsubsection{Function: {\tt get\_ranges}}

This function converts any of the four input types for {\tt genColType} to a vector with the relevant values to keep on a given index.  This is then used in {\tt setindex!} principally.

\begin{lstlisting}[language=julia]
function get_ranges(C::Tuple,a::genColType...)
  ap = Array{genColType,1}(undef,length(a))
  for y = 1:length(ap)
    if typeof(a[y]) <: Colon
      ap[y] = 1:C[y]
    elseif typeof(a[y]) <: Integer
      ap[y] = a[y]:a[y]
    elseif typeof(a[y]) <: AbstractArray
      ap[y] = 1:length(a[y])
    else
      ap[y] = a[y]
    end
  end
  return ap
end
\end{lstlisting}

The generation of the ranges is somewhat more costly in terms of allocations for a temporary object, but the overall cost for this operation is very low inside of a given algorithm.

\subsubsection{Function: {\tt makepos} \& {\tt position\_incrementer!}}

When writing loops in general, there are two basic strategies if an array of numbers need to be iterated over.  One option is to use the modulus operation to convert an index to a position (see the conversions between integers and positions in the {\tt pos2ind} and {\tt ind2pos} functions).  The second option is to explicitly increment a position vector, starting at zero, and continue until the maximum length is reached.  The functions {\tt makepos} and {\tt position\_incrementer!} generate the initial position and then increment it.  The second function is mainly used inside of a function.

\begin{lstlisting}[language=julia]
function makepos(ninds::P;zero::P=0) where P <: Integer
  pos = Array{P,1}(undef,ninds)
  pos[1] = zero
  one = zero + 1
  for g = 2:ninds
    pos[g] = one
  end
  return pos
end
export makepos

function position_incrementer!(pos::Array{G,1},sizes::Union{Array{G,1},Tuple}) 
				where G <: intType
  w = 1
  @inbounds pos[w] += 1
  @inbounds while w < length(sizes) && pos[w] > sizes[w]
    @inbounds pos[w] = 1
    w += 1
    @inbounds pos[w] += 1
  end
  nothing
end
export position_incrementer!
\end{lstlisting}

The use of the {\tt makepos} function deserves some discussion.  One can choose to initialize the vector as a 1-indexed vector or a 0-indexed vector by choosing either {\tt zero=0} or {\tt zero=-1}, respectively.

\subsection{Elementary operations}

These functions will closely resemble julia's interface, but will be necessary to define in other implementations.

\subsubsection{Function: {\tt copy}}

The most robust copy function in julia is the {\tt deepcopy} function, but using a {\tt deepcopy} is inherently type unstable, meaning that the compiler can not always resolve the output type from the function.  One can use this function whenever the need to copy an array is required, but defining a suitable {\tt copy} function instead will make julia's other function and keep the code type stable.  The function {\tt deepcopy} works immediately on the previously defined types.

\begin{lstlisting}[language=julia]
import Base.copy
function copy(A::tens{W}) where {W <: Number}
  return tens{W}(A.size,copy(A.T))
end
\end{lstlisting}

The copy function must be careful to leave no possible way for changes in the original tensor to affect the new tensor. So, the main item that must be addressed here is the {\tt T} field.  For example, if the {\tt copy} operation is not applied on this tensor, then conjugating the original elements will conjugate both tensors, even if only one of the tensors is to be conjugated!  The reason that this happens because simple definitions of variables are not inherently copies in julia, instead they are passed by reference.  So the new object only stores a pointer that references the original data.  If the original data is changed, so will the new object.  The tuple must remain unchanged, so it is not necessary to copy.

When making a previously defined function, it is often the case that trial and error will determine which quantities need to be copied and which do not.  For safety, one can simply copy every field, but there is a very small savings that can be found with careful copying.

\subsubsection{Function: {\tt length}}

This returns the number of elements in the tensor in julia, so it will perform the same operation here.

\begin{lstlisting}[language=julia]
import Base.length
function length(M::denstens)
  return length(M.T)
end
\end{lstlisting}

\subsubsection{Function: {\tt size}}

This returns the size of a given index on a tensor.  If the full tensor is input, then the Tuple representing the size of the specified index is returned.

\begin{lstlisting}[language=julia]
import LinearAlgebra.size
import Base.size
function size(A::denstens)
  return A.size
end

function size(A::denstens,i::Integer)
  return A.size[i]
end
\end{lstlisting}

\subsubsection{Function: {\tt sum}}

This returns the sum of the elements of a given index on a tensor. This can be accomplished by simply summing up the values in the {\tt T} field (stored values) of the tensor.

\begin{lstlisting}[language=julia]
import Base.sum
function sum(A::denstens)
  return sum(A.T)
end
\end{lstlisting}

\subsubsection{Function: {\tt norm}}

This returns the $L^1$-norm (Froebenius norm) of the elements of a given index on a tensor.

\begin{lstlisting}[language=julia]
import LinearAlgebra.norm
function norm(A::denstens)
  return norm(A.T)
end
export norm
\end{lstlisting}

\subsubsection{Function: {\tt conj} \& {\tt conj!}}

To conjugate all elements of a tensor, the {\tt conj} function can be used.  This will return a copy of the original tensor.  Alternatively, one can conjugate in-place with {\tt conj!}. Both of these functions are overloaded from julia's internal implementations.

\begin{lstlisting}[language=julia]
import LinearAlgebra.conj!
function conj!(currtens::denstens)
  LinearAlgebra.conj!(currtens.T)
  return currtens
end

import LinearAlgebra.conj
function conj(M::tens{W}) where W <: Number
  newT = LinearAlgebra.conj(M.T)
  return tens{W}(M.size,newT)
end
\end{lstlisting}

\subsubsection{Function: {\tt ndims} (tensor rank)}

This function returns the rank of the tensor, overloading julia's internal function.

\begin{lstlisting}[language=julia]
import LinearAlgebra.ndims
function ndims(A::denstens)
  return length(A.size)
end
\end{lstlisting}

\subsubsection{Function: {\tt lastindex}}

This function is necessary to determine the last element of a vector.  Applied to the {\tt denstens} type, the return value is the last size.  This is used in julia when {\tt end} is called on a particular tensor ({\it i.e.}, {\tt 1:end}).

\begin{lstlisting}[language=julia]
import Base.lastindex
function lastindex(M::denstens, i::Integer)
  return M.size[i]
end
\end{lstlisting}

\subsubsection{Function: {\tt eltype}}

This returns the element type in the dense tensor.

\begin{lstlisting}[language=julia]
import Base.eltype
function eltype(A::tens{W}) where W <: Number
  return W
end
\end{lstlisting}

\subsubsection{Function: {\tt elnumtype}}

This returns the element type in the dense tensor.  This is the same behavior as {\tt eltype}, but this function will have a new meaning when dealing with MPSs and MPOs.

\begin{lstlisting}[language=julia]
function elnumtype(A::tens{W}) where W <: Number
  return eltype(A)
end
export elnumtype
\end{lstlisting}

\subsubsection{Function: {\tt getindex} \& {\tt getindex!}}

To obtain a single element of the tensor, or to extract a subset of the values in a tensor, the function {\tt getindex} (and its in-place variation {\tt getindex!}) can be called.  The function {\tt getindex} overloads julia's internal function for the {\tt denstens} type and the in-place version performs the same with a minor cost savings.

\begin{lstlisting}[language=julia]
import Base.getindex
function getindex(M::denstens, a::genColType...)
  return getindex!(M,a...)
end

function getindex(C::tens{W}, a::G...)where W <: Number where G <: Array{Bool,1}
  M = makeArray(C)
  return tens{W}(M[a...])
end

function getindex!(C::AbstractArray, a::genColType...)
  return getindex(C, a...)
end

function getindex!(C::tens{W}, a::genColType...) where W <: Number
  i = 0
  allintegers = true
  while allintegers && i < length(a)
    i += 1
    allintegers = typeof(a[i]) <: Integer
  end
  if allintegers
    val = pos2ind(a,C.size)
    return C.T[val]
  else
    dC = makeArray(C)[a...]
    return tens{W}(dC)
  end
end
export getindex!
\end{lstlisting}

Much of the work of extract the elements relies on julia's internal system which uses strided arrays.  DMRjulia may implement a version of this in the future.

Note that there is a differentiation between inputting a value of 1 for a given entry of {\tt a} and the entry {\tt 1:1} yield very different results.  For the integer input of 1, a single value is returned and the index is truncated. If the range {\tt 1:1} is input, then the index is not truncated and in general the array properties of the tensor are kept.  In other words, the range implies that a range should be output while the integer signifies that the index should be excluded from the final tensor.

\subsubsection{Function: {\tt searchindex}}

The function {\tt getindex} has many extra components in it that aren't necessary if a single set of integers must be found in a tensor. The function {\tt searchindex} is a faster method that is sometimes used in the library to search for a single element of a tensor. The basic strategy is to generate the index position (a single integer) of where the element should be in the dense tensor and then extract that element. 

\begin{lstlisting}[language=julia]
function searchindex(C::denstens,a::intType...)
  if length(C.T) == 0
    outnum = 0
  elseif length(C.size) == 0
    outnum = C.T[1]
  else
    veca = ntuple(i->a[i],length(a))
    w = pos2ind(veca,C.size)
    outnum = C.T[w]
  end
  return outnum
end
export searchindex
\end{lstlisting}

Note that generating a single integer {\tt w} inside of the for loop does not create much overhead for the julia language. So, the in-place version of {\tt pos2ind} is not defined and used here. This function might not be useful in lower level implementations of the tensor networks, but it can be useful here to obtain a single element at low cost from a tensor, returned as a scalar, in a unified way between all tensor types.

\subsubsection{Function: {\tt setindex} \& {\tt setindex!}}

In many cases, the elements of a tensor should be set.  This is not strictly required for a generic tensor network computation, and is not a feature in some libraries, but in some cases it is found to be useful here.  The function reduces to the julia function in line with the treatment for {\tt getindex}.

\begin{lstlisting}[language=julia]
import Base.setindex!
function setindex!(B::denstens,A::G,a::genColType...) 
				where G <: Union{denstens,AbstractArray}
  indexranges = get_ranges(B.size,a...)
  nterms = prod(newsize)

  Asize = size(A)
  Bsize = size(B)
  for z = 1:nterms
    ind2pos!(pos,z,Asize)
    @simd for i = 1:length(indexranges)
      @inbounds s = pos[i]
      @inbounds pos[i] = indexranges[i][s]
    end
    x = pos2ind(pos,Bsize)
    @inbounds B.T[x] = A.T[z]
  end
  nothing
end

function setindex!(B::tens{W},A::W,a::Integer...) 
				where W <: Number
  @inbounds index = a[end]-1
  @simd for q = length(a)-1:-1:1
    @inbounds index *= size(B,q)
    @inbounds index += a[q]-1
  end
  @inbounds B.T[index+1] = A
  nothing
end
\end{lstlisting}

Each of the values to replace is scanned over and copied from the input.

The function is parallelized over all the inputs that must be adjusted and the $z$ value that indexes the number of terms to be replaced in the tensor is computed using the relatively expensive modulus operation in {\tt pos2ind!}.  This is mostly offset by the parallelization for typical applications.  Because in the worst case of a dense square matrix this would scale as $O(m^2)$, the choice was made to use parallelization instead of the incrementor position method from {\tt position\_incrementer!}.

\subsection{Linear Algebra}

Basic linear algebra operations on tensors are a fundamental function that must be defined for many algorithms. Many of these can be written into a single function {\tt tensorcombination} that handles more simple algebraic operations.

\subsubsection{Function: {\tt loadM!}}

{\tt loadM!} is a simple copy operation to the left input matrix from the right input matrix. Assumes same element type and useful for ensuring compiler efficiency.

\begin{lstlisting}[language=julia]
function loadM!(output::Array{W,N},input::Array{W,N}) where {N, W <: Number}
  @simd for x = 1:length(output)
    @inbounds output[x] = input[x]
  end
  nothing
end
\end{lstlisting}

Otherwise, julia may assume that a different type is being copied, thus forcing a conversion and an extra allocation to be thrown.

\subsubsection{Function: {\tt tensorcombination!}}

This function forms the basis of all algebraic manipulations of {\tt denstens} going forward.  This will create a new linear combination of all input tensors with coefficients supplied by alpha.  For example, the output $K$ is a linear combination of input tensors
\begin{equation}\label{linearcombination}
K=\sum_i\alpha_iM_i
\end{equation}
where the set of $M_i$ tensors was input into the function with coefficients $\alpha_i$. 

\begin{lstlisting}[language=julia]
function tensorcombination(M::denstens...;alpha::Tuple=ntuple(i->1,length(M)),
				fct::Function=*)
  alphamult = prod(alpha)
  Mmult = prod(i->M[i].T[1],1:length(M))
  outType = typeof(alphamult * Mmult)
  newTensor = tens{outType}(M[1].size,zeros(outType,length(M[1])))
  for i = 1:length(M[1])
    @inbounds newTensor.T[i] = 0
    @simd for k = 1:length(M)
      @inbounds newTensor.T[i] += fct(M[k].T[i],alpha[k])
    end
  end
  return newTensor
end

function tensorcombination(M::AbstractArray...;
			alpha::Tuple=ntuple(i->1,length(M)),fct::Function=*)
  return sum(k->fct(M[k],alpha[k]),1:length(M))
end

function tensorcombination(alpha::Tuple,M::P...;fct::Function=*) 
				where P <: TensType
  return tensorcombination(M...,alpha=alpha,fct=fct)
end
export tensorcombination

function tensorcombination!(M::denstens...;alpha::Tuple=ntuple(i->1,length(M)),
				fct::Function=*) where W <: Number
  for i = 1:length(M[1])
    x = 0
    @simd for k = 1:length(M)
      @inbounds x += fct(M[k].T[i],alpha[k])
    end
    M[1].T[i] = x
  end
  return M[1]
end

function tensorcombination!(M::AbstractArray...;
			alpha::Tuple=ntuple(i->1,length(M)),fct::Function=*)
  return tensorcombination(M...,alpha=alpha,fct=fct)
end

function tensorcombination!(alpha::Tuple,M::P...;fct::Function=*) 
						where P <: TensType
  return tensorcombination!(M...,alpha=alpha,fct=fct)
end
export tensorcombination!
\end{lstlisting}

The inputs to this function are allowed to best done with tuples, but this can be input with a vector as well. As per usual, the exclamation point (!) signifies that the leading tensor is modified while the rest are not.  If the exclamation point is not included, then the first tensor is copied and then modified.  The usefulness of this function is that it can reduce code for the other functions in this section.

Note that the combination function {\tt fct} can be modified.  This means that one could replace it with, for example, division ({\tt /}) or any other operation defined for scalars and tensors or even to implement a square root function. The default is Eq.~\eqref{linearcombination}.

\subsubsection{Function: {\tt mult!}}

The multiplication of a scalar with a tensor can be accomplished by simply calling the {\tt tensorcombination} function with a single input.

\begin{lstlisting}[language=julia]
function mult!(M::P, num::Number) where P <: TensType
  return tensorcombination!(M,alpha=(num,))
end

function mult!(num::Number,M::P) where P <: TensType
  return mult!(M,num)
end
export mult!
\end{lstlisting}

\subsubsection{Function: {\tt add!}}

The addition of two tensors {\tt A} and {\tt B} where {\tt B} receives a scalar multiplicative factor is defined with the {\tt add!} function.

\begin{lstlisting}[language=julia]
function add!(A::P, B::P, num::Number) where P <: TensType
  return tensorcombination!((1,num),A,B)
end

function add!(A::P, B::P) where P <: TensType
  return add!(A,B,1)
end
export add!
\end{lstlisting}

\subsubsection{Function: {\tt sub!}}

The subtraction of two tensors {\tt A} and {\tt B} where {\tt B} receives a scalar multiplicative factor is defined with the {\tt sub!} function. The factor {\tt mult} is simply negated to get the subtraction and then the {\tt add!} function is called.

\begin{lstlisting}[language=julia]
function sub!(A::P,B::P,mult::Number) where P <: TensType
  return add!(A,B,-mult)
end

function sub!(A::P,B::P) where P <: TensType
  return add!(A,B,-1)
end
export sub!
\end{lstlisting}

\subsubsection{Function: {\tt div!}}

Division of a tensor by a scalar is identical to multiplication by the inverse of that scalar value.  At least to the typically used double precision, this is accurate enough, so the {\tt div!} function is defined as

\begin{lstlisting}[language=julia]
function div!(M::P, num::Number) where P <: TensType
  return tensorcombination!(M,alpha=(num,),fct=/)
end
export div!
\end{lstlisting}

\subsubsection{Functions: {\tt +,-,*,/}}

The basic functions of addition, subtraction, multiplication, and division can be defined to give a tensor that is copied from the original.

\begin{lstlisting}[language=julia]
import LinearAlgebra.+
function +(A::P, B::P) where P <: TensType
  return tensorcombination(A,B)
end

import LinearAlgebra.-
function -(A::P, B::P) where P <: TensType
  return tensorcombination((1,-1),A,B)
end

import Base.*
function *(num::Number, M::P) where P <: TensType
  return tensorcombination(M,alpha=(num,))
end

function *(M::P, num::Number) where P <: TensType
  return num * M
end

import LinearAlgebra./
function /(M::P, num::Number) where P <: TensType
  return tensorcombination(M,alpha=(num,),fct=/)
end
\end{lstlisting}

Each function calls the {\tt tensorcombination} function and returns a completely new tensor, so no in-place functions are called here.  To perform those operations, the {\tt add!}, {\tt sub!}, {\tt mult!}, and {\tt div!} functions can be used.

\subsubsection{Functions: {\tt sqrt!} \& {\tt sqrt}}

The square root can be defined from {\tt tensorcombination} just as before with the other functions.  The output is either manipulated in-place on the first tensor (!) or not.

\begin{lstlisting}[language=julia]
function sqrt!(M::TensType)
  return tensorcombination!(M,alpha=(0.5,),fct=^)
end
export sqrt!

import Base.sqrt
function sqrt(M::TensType)
  return tensorcombination(M,alpha=(0.5,),fct=^)
end
\end{lstlisting}

\subsubsection{Functions: {\tt inverse\_element} \& {\tt invmat}}

The inverse of an element can be taken in a tensor.

\begin{lstlisting}[language=julia]
function inverse_element(x::Number,zeronum::Real)
  return abs(x) > zeronum ? 1/x : 0.
end
\end{lstlisting}

This is most commonly performed on the diagonal elements of the $D$ matrix returned from a singular value decomposition. This can appear in some algorithms.

The function {\tt inverse\_element} can be used to find the inverse of a diagonal matrix with {\tt invmat}.

\begin{lstlisting}[language=julia]
function invmat!(M::denstens;zeronum::Float64=1E-16)
  return tensorcombination!(M,alpha=(zeronum,),fct=inverse_element)
end

function invmat!(M::AbstractArray;zeronum::Float64=1E-16)
  for a = 1:length(M)
    M[a,a] = abs(M[a,a]) > zeronum ? 1/M[a,a] : 0.
  end
  return M
end
export invmat!

function invmat(M::denstens;zeronum::Float64=1E-16)
  return tensorcombination(M,alpha=(zeronum,),fct=inverse_element)
end

function invmat(M::AbstractArray;zeronum::Float64=1E-16)
  outM = zeros(eltype(M),size(M))
  for a = 1:length(M)
    outM = abs(M[a,a]) > zeronum ? 1/M[a,a] : 0.
  end
  return outM
end
export invmat
\end{lstlisting}

The same function is also defined for julia's base {\tt Array} class. This is in case any algorithm must be run with the {\tt Array} struct.

\subsubsection{Functions: {\tt exp}}

Exponentiating a matrix is an internal function to julia.  The startegy is to take a matrix, $\hat G$, and perform an eigenvalue decomposition.  The diagonal matrix has each diagonal element exponentiated.  The unitary matrices remain unchanged as can be seen by an explicit Taylor series expansion
\begin{equation}
\exp(\hat G)=\hat U\exp(\hat S)\hat U^\dagger
\end{equation}
where $\exp(\hat S)$ means to take the elements of the diagonal $\hat S$ matrix (found by eigenvalue decomposition) and exponentiating each.

\begin{lstlisting}[language=julia]
import Base.exp
function exp(A::tens{W}) where W <: Number
  X = reshape(A.T,A.size)
  return tens(W,exp(X))
end
\end{lstlisting}

\subsection{Tensor operations}

The operations covered here are for a single tensor.  These are two of the four basic operations identified in Ref.~\onlinecite{bakerCJP21}.

\subsubsection{Function: {\tt reshape!}}\label{reshapefunction}

The first basic operation from Ref.~\onlinecite{bakerCJP21,*baker2019m} is the reshaping of a dense tensor. The action is very simple, the {\tt size} field must be replaced by a new tuple.  The number of elements does not change, so the elements of the new tuple must multiply to the same number as the original tuple.

There are several variations on the {\tt reshape} function that are included here.  One is to perform an in-place reshape, {\tt reshape!}.  This does a shallow copy on the elements of the tensor network ({\tt T} field) and simply replaces the {\tt size}.  A version of this function is also defined for julia's array types.

\begin{lstlisting}[language=julia]
function reshape!(M::tens{W}, S::NTuple{N,intType};
		merge::Bool=false) where {N, W <: Number}
  M.size = S
  return M
end

function reshape!(M::tens{W}, S::intType...;merge::Bool=false) where W <: Number
  return reshape!(M,S)
end

function reshape!(M::tens{W}, S::Array{Array{P,1},1};merge::Bool=false) 
			where {W <: Number, P <: Integer}
  newsize = ntuple(a->prod(b->size(M,b),S[a]),length(S))
  order = vcat(S...)
  pM = issorted(order) ? M : permutedims!(M,order)
  return reshape!(M,newsize)
end
export reshape!
\end{lstlisting}

At this time, julia does not have an established system for disabling errors in functions, and the option taken here is for fast code over error-proof code.  If there was an option, then it would be added here, but instead we rely on using julia's native tensor format to check if the reshapes are done correctly, then convert to the dense tensor type.

Sometimes it is useful to input the reshape in terms of the bare indices on the tensor.  This can be useful for bookkeeping and avoiding the need to type out lengthy expressions that compute the new size of a reshaped tensor.  The example syntax is {\tt reshape(A,[[1,2],[3]])} which reshapes the first two indices together of a rank-3 tensor.  The equivalent in the other expression would look something like {\tt reshape(A,size(A,1)*size(A,2),size(A,3))} generically, although other versions can be defined.

\subsubsection{Function: {\tt reshape}}

Depending on the use of the {\tt reshape} function, sometimes the function will need to be reshaped with a copy operation applied to the input tensor.  This is accomplished with the following function.

\begin{lstlisting}[language=julia]
import Base.reshape
function reshape(M::tens{W}, S::intType...;merge::Bool=false) where W <: Number
  newMsize = S
  newM = tens{W}(newMsize,M.T)
  return reshape!(newM,S)
end

function reshape(M::tens{W}, S::Array{Array{P,1},1};merge::Bool=false) 
			where {W <: Number, P <: intType}
  return reshape!(copy(M),S)
end
\end{lstlisting}

\subsubsection{Function: {\tt unreshape} \& {\tt unreshape!}}

This function was primarily introduced for documentation purposes.  It is easier to read a reshape that restores the original shape if it goes by another name.  Here, this merely acts as a {\tt reshape} with a different name.

\begin{lstlisting}[language=julia]
function unreshape!(M::AbstractArray, S::intType...;merge::Bool=false) 
				where W <: Number
  return reshape(M,S...)
end

function unreshape!(M::tens{W}, S::intType...;merge::Bool=false) 
				where W <: Number
  return reshape!(M,S)
end
export unreshape!

function unreshape(M::AbstractArray, S::intType...;merge::Bool=false) 
				where W <: Number
  return reshape(M,S...)
end

function unreshape(M::tens{W}, S::intType...;merge::Bool=false) 
				where W <: Number
  newM = tens{W}(M.size,M.T)
  return unreshape!(newM,S...)
end
export unreshape
\end{lstlisting}

There is a useful modification of this function for the quantum number case, and that was the second reason for defining this function.

\subsubsection{Function: {\tt permutedims!}}

In addition to {\tt permutedims}, there is also the possibility to perform this operations in place.  To do this, the {\tt permutedims!} operation from julia is used, similarly to the discussion around {\tt permutedims} (Sec.~\ref{permutedimsfunction}). 

\begin{lstlisting}[language=julia]
import Base.permutedims!
function permutedims!(M::tens{W}, vec::Array{P,1}) 
				where {W <: Number, P <: intType}
  return permutedims!(M,(vec...,),M.size...)
end

function permutedims!(M::tens{W}, vec::NTuple{N,intType}) where {N,W <: Number}
  return permutedims!(M,vec,M.size...)
end

function permutedims!(M::AbstractArray, 
			vec::NTuple{N,intType}) where {N, W <: Number}
  return permutedims(M,vec)
end

function permutedims!(M::tens{W}, vec::NTuple{N,intType},
			x::intType...) where {N,W <: Number}
  rM = reshape(M.T,x)
  xM = permutedims(rM, vec)
  out = tens(W,xM)
  return out
end
\end{lstlisting}

There is only a small difference when using the two functions, but it can be convenient to use this function when creating new code instead of designating a new tensor each time.

\subsubsection{Function: {\tt permutedims}}\label{permutedimsfunction}

Permuting the dimensions of a tensor is a basic operation that is probably defined in every language where tensor network methods will be used.  It is a standard function in computer science, and while developing DMRjulia, it was questioned whether to recode this function from scratch.  But the existing functions were so efficient and involved clever algorithms from computer science, that this function will be left to existing libraries, so that the focus here can be on the operations specifically necessary for tensor networks.  

\begin{lstlisting}[language=julia]
import Base.permutedims
function permutedims(M::tens{W}, vec::Array{P,1}) 
				where {W <: Number, P <: intType}
  return permutedims!(M,(vec...,))
end
\end{lstlisting}

\subsubsection{Function: {\tt joinindex} \& {\tt joinindex!}}

It is sometimes the case that two tensors must be joined along a given index.  For example, if two rank-3 tensors are to be joined along the third index of each tensor (with all other tensors kept the same), then the function {\tt joinindex} (or in-place {\tt joinindex!}) can be called as {\tt joinindex(A,B,[1,3])} to accomplish this.  This is used mostly for the single site implementation of DMRG, although it can have other uses as well.

\begin{lstlisting}[language=julia]
function joinindex!(bareinds::intvecType,A::Array{R,N},
			B::Array{S,N})::Array{typeof(R(1)*S(1)),N} 
			where {R <: Number, S <: Number, N}
  inds = convIn(bareinds)
  
  nA = ndims(A)
  nB = nA
  finalsize = [size(A,i) for i = 1:nA]
  @simd for i = 1:length(inds)
    @inbounds a = inds[i]
    @inbounds finalsize[a] += size(B,a)
  end
  thistype = typeof(S(1)*R(1))
  if length(inds) > 1
    newTensor = zeros(thistype,finalsize...)
  else
    dim = length(finalsize)
    newTensor = Array{thistype,dim}(undef,finalsize...)
  end

  notinds = findnotcons(ndims(A),(inds...,))
  axesout = Array{UnitRange{intType},1}(undef,nA)
  @simd for i = 1:length(notinds)
    @inbounds a = notinds[i]
    @inbounds axesout[a] = 1:finalsize[a]
  end
  @simd for i = 1:length(inds)
    @inbounds b = inds[i]
    @inbounds start = size(A,b)+1
    @inbounds stop = finalsize[b]
    @inbounds axesout[b] = start:stop
  end

  axesA = ntuple(a->1:size(A,a),nA)
  @inbounds newTensor[axesA...] = A
  @inbounds newTensor[axesout...] = B

  return newTensor
end

function joinindex(inds::intvecType,A::W,B::R...) 
			where {W <: TensType, R <: TensType}
  if typeof(A) <: densTensType
    C = A
  else
    C = copy(A)
  end
  return joinindex!(inds,C,B...)
end

function joinindex(A::W,B::R,inds::intvecType) 
			where {W <: TensType, R <: TensType}
  mA,mB = checkType(A,B)
  return joinindex(inds,mA,mB)
end
export joinindex

function joinindex!(A::W,B::R,inds::intvecType) 
			where {W <: TensType, R <: TensType}
  mA,mB = checkType(A,B)
  return joinindex!(inds,mA,mB)
end
export joinindex!
\end{lstlisting}

\subsection{Display}

\subsubsection{Function: {\tt showTens}}

The function {\tt showTens} outputs the most relevant information for an input {\tt denstens}.  The size is displayed, and the tensor is output to {\tt show} digits (default 4).

\begin{lstlisting}[language=julia]
function showTens(M::denstens;show::intType = 4)
  println("printing regular tensor of type: ", typeof(M))
  println("size = ", M.size)
  maxshow = min(show, size(M.T, 1))
  maxBool = show < size(M.T, 1)
  println("T = ", M.T[1:maxshow], maxBool ? "..." : "")
  nothing
end
export showTens
\end{lstlisting}

\subsubsection{Function: {\tt print}}

This prints out the relevant information from {\tt showTens}

\begin{lstlisting}[language=julia]
import Base.print
function print(A::denstens...;show::intType = 4)
  showTens(A, show = show)
  nothing
end
\end{lstlisting}

\subsubsection{Function: {\tt println}}

This is the print operation but an extra return is entered to start a new line.  This is the most commonly used version in {\tt julia}.

\begin{lstlisting}[language=julia]
import Base.println
function println(A::denstens;show::intType = 4)
  showTens(A, show = show)
  print("\n")
  nothing
end
\end{lstlisting}

\section{Module: {\tt Opt.jl}}

There are several operators that are frequently used for several lattice types.  Normally, one would have to define some special properties with this, but here the julia arrays can be used, simply.


The three sets of operators will be defined for spins, Hubbard models (fermions), and $t-J$ models.

\subsubsection{Function: {\tt spinOps.jl}}

The spin operators that are generated are the standard operators for a given given spin system of dimension $s$. The standard relations are used to determine the matrix elements. The Pauli matrices are defined as \cite{townsend2000modern}
\begin{equation}
\hat S^z_{a,b}=(s+1-a)\delta_{a,b},\quad \hat S^{+}_{m,m+1}=\frac{\delta_{m,m+1}}2\sqrt{s(s+1)-m(m+1)},\quad \hat S^-=\left(\hat S^+\right)^\dagger
\end{equation}
where $a$ and $b$ index the rows and columns of the matrices.

The letter {\tt O} is reserved as a zero element.  It is only full of zeros, simply.  The matrix {\tt Id} is used as an identity operation.  These are useful when constructing MPOs manually.  The function was designed for integer and half-integer spin of any magnitude.

\begin{lstlisting}[language=julia]
function spinOps(;s=0.5)
  states = convert(Int64,2*s+1) #number of quantum states

  O = zeros(Float64,states,states) #zero matrix
  Id = copy(O) + LinearAlgebra.I #identity matrix
  oz = copy(O) # z operator
  op = copy(O) # raising operator
  for (q,m) in enumerate(s:-1:-s) #counts from m to -m (all states)
    oz[q,q] = m
    if m+1 <= s
      op[q-1,q] = sqrt(s*(s+1)-m*(m+1))
    end
  end
  om = Array(op') # lowering operator
  ox = (op+om)/2 #x matrix
  oy = (op-om)/(2*im) #y matrix

  return ox,oy,oz,op,om,O,Id
end

function spinOps(a::Float64)
  return spinOps(s=a)
end
export spinOps
\end{lstlisting}

Note the order output from the function and how it must appear when loading these operators into a program.   If another operator is required, such as $\hat S^z\cdot \hat S^z$, then the syntax to generate this is simply {\tt Sz*Sz}.

\subsubsection{Function: {\tt fermionOps.jl}}

The code to generate the fermion operators under this construction is \cite{hubbard1963electron,senechal2008introduction}

\begin{lstlisting}[language=julia]
function fermionOps()
  states = 4 #fock space size
  O = zeros(Float64,states,states) #zero matrix
  Id = copy(O)+LinearAlgebra.I #identity

  Cup = copy(O) #annihilate (up)
  Cup[1,2] = 1.
  Cup[3,4] = 1.

  Cdn = copy(O) #annihilate (down)
  Cdn[1,3] = 1.
  Cdn[2,4] = -1.

  Nup = Cup' * Cup #density (up)
  Ndn = Cdn' * Cdn #density (down)
  Ndens = Nup + Ndn #density (up + down)

  F = copy(Id) #Jordan-Wigner string operator
  F[2,2] *= -1.
  F[3,3] *= -1.

  return Cup,Cdn,Nup,Ndn,Ndens,F,O,Id
end
export fermionOps
\end{lstlisting}

\subsubsection{Function: {\tt tJOps.jl}}

The $t-J$ model is the same as the Hubbard model, but with double occupancy projected out. So, the operators from the Hubbard model are created and then only the first 3 states are kept.

\begin{lstlisting}[language=julia]
function tJOps()
  #many of the Hubbard operators can be truncated
  Cup,Cdn,Nup,Ndn,Ndens,F,O,Id = fermionOps()
  states = 3 #fock space size
  s = states
  Cup = Cup[1:s,1:s]
  Cdn = Cdn[1:s,1:s]
  Nup = Nup[1:s,1:s]
  Ndn = Ndn[1:s,1:s]
  Ndens = Ndens[1:s,1:s]
  F = F[1:s,1:s]
  O = O[1:s,1:s]
  Id = Id[1:s,1:s]

  Sz = copy(O) #z-spin operator
  Sz[2,2] = 0.5
  Sz[3,3] = -0.5

  Sp = copy(O) #spin raising operator
  Sp[3,2] = 1.
  Sm = Array(Sp') #spin lowering operator

  return Cup,Cdn,Nup,Ndn,Ndens,F,Sz,Sp,Sm,O,Id
end
export tJOps
\end{lstlisting}

The spin operators are sometimes used in the context of $t-J$ models and are implemented here according to the relationship \cite{spalek2007tj}
\begin{equation}
\mathbf{S}=\langle\Psi|\hat c^\dagger_{i,\sigma}\vec{\sigma}_{\sigma\sigma'}\hat c_{i,\sigma'}|\Psi\rangle
\end{equation}
To recover the matrix elements as was done for the fermionic operators to see how the matrix elements are given in the code.

\section{Module: {\tt contractions.jl}}\label{contractionsmodule}

As the first example of one of the core operations in a tensor network, it is now shown how to code an efficient conversion algorithm to contract two tensors by using a matrix multiplication algorithm from a known library.  The implementation may change for different languages that this is written in, but the basic concepts will remain the same. 

\subsection{Conversion of tensors to and evaluation of matrix multiplication}

\subsubsection{Function: {\tt libmult!}}

This is the core of the contraction algorithm.  A function from the BLAS library is used to compute the relevant contraction.  One feature that must be determined before the BLAS function is whether the types of all inputs is correct.  That is done as a pre-processing step before activating the BLAS library.

\begin{lstlisting}[language=julia]
function libmult(C::Array{W,2},D::Array{X,2},tempZ::Array{P,2}...;
		alpha::Number=1.,beta::Number=1.) 
		where {W <: Number, X <: Number, P <: Number}
  if length(tempZ) == 0
    if (eltype(C) == eltype(D)) && !(eltype(C) <: Integer)
      temp_alpha = typeof(alpha)<:eltype(C) ?  alpha : convert(eltype(C),alpha)
      retMat = LinearAlgebra.BLAS.gemm('N','N',temp_alpha,C,D)
    else
      retMat = alpha * C * D
    end
  else
    if (eltype(C) == eltype(D) == eltype(Z[1])) && !(eltype(C) <: Integer)
      temp_alpha = convert(eltype(C),alpha)
      temp_beta = convert(eltype(C),beta)
      LinearAlgebra.BLAS.gemm!('N','N',temp_alpha,C,D,temp_beta,tempZ[1])
      retMat = tempZ[1]
    else
      retMat = alpha * C * D + beta * tempZ[1]
    end
  end
  return retMat
end
export libmult
\end{lstlisting}

The innermost matrix multiplication is often crucial to define.  Even a small inefficiency can create a large time lag on the whole algorithm. At the present time, julia's BLAS library does not have all the features that could be implemented here.  For example, changing the `N' inputs (normal matrix) to `T' for transpose cases almost not change in the computational speed even for very large matrices. If at a future time there are additional functionalities that are input into julia, then this is the function that must change to accommodate them.

A graphics processing unit (GPU) implementation of this function is still undergoing testing as of this version, but this is the place where such an implementation would go.

\subsubsection{Function: {\tt matrixequivalent} \& {\tt matrixequiavlent!}}

The function {\tt matrixequivalent} generates the matrix equivalent of a given input tensor.  This can also happen in place.  The main reason for defining this function is that it is slightly easier for the compiler to handle the input if the types are separated.

\begin{lstlisting}[language=julia]
function matrixequiv(X::densTensType,Lsize::Integer,Rsize::Integer)
  return matrixequiv!(copy(X),Lsize,Rsize)
end

function matrixequiv!(X::AbstractArray,Lsize::Integer,Rsize::Integer)
  return reshape!(X,Lsize,Rsize)
end

function matrixequiv!(X::denstens,Lsize::Integer,Rsize::Integer)
  return reshape!(X.T,Lsize,Rsize)
end
\end{lstlisting}

The definition of these functions might seem to be an over-use of the function structure here, but julia often optimizes code at the function barriers to ensure type stability. The functions are kept here largely in case a future compiler version is aided by these functions. The same is true for the next function.

\subsubsection{Function: {\tt permutedims\_2matrix!}}

The function {\tt permutedims\_2matrix!} generates the matrix equivalent of a given input tensor after first permuting some of the indices.  

\begin{lstlisting}[language=julia]
function permutedims_2matrix!(X::AbstractArray,vec::Tuple,
			Lsize::Integer,Rsize::Integer)
  xM = permutedims(X, vec)
  return reshape!(xM,Lsize,Rsize)
end

function permutedims_2matrix!(X::denstens,vec::Tuple,
			Lsize::Integer,Rsize::Integer)
  return permutedims_2matrix!(makeArray(X),vec,Lsize,Rsize)
end
\end{lstlisting}

\subsubsection{Function: {\tt prepareT}}

The function {\tt prepareT} will make the matrix equivalent of the tensor based on the conjugation and contracted indices.  The return values are then multiplied together and the result is then unreshaped when passed back to {\tt maincontractor}.

\begin{lstlisting}[language=julia]
function prepareT(A::densTensType,Lvec::Tuple,Rvec::Tuple,conjvar::Bool)
  vec = (Lvec...,Rvec...)

  Lsize = length(Lvec) > 0 ? prod(a->size(A,a),Lvec) : 1
  Rsize = length(Rvec) > 0 ? prod(a->size(A,a),Rvec) : 1

  issort = true
  a = 1
  while issort && a < length(Lvec)
    a += 1
    issort = Lvec[a-1] < Lvec[a]
  end
  if length(Rvec) > 0 && length(Lvec) > 0
    issort = issort && Lvec[end] < Rvec[1]
    a = 1
  end
  while issort && a < length(Rvec)
    a += 1
    issort = Rvec[a-1] < Rvec[a]
  end

  if issort
    if eltype(A) <: Complex && conjvar
      out = matrixequiv(A,Lsize,Rsize)
    else
      out = matrixequiv!(A,Lsize,Rsize)
    end
  else
    out = permutedims_2matrix!(A,vec,Lsize,Rsize)
  end
  if conjvar
    conj!(out)
  end
  return out
end
\end{lstlisting}

This function takes care not compute a permutation or conjugation when it is not necessary ({\it i.e.}, indices are already ordered or element type is real).  It will only copy the input when it must.  This is the function where it is convenient to conjugate the input tensor because the matrix equivalent can be regarded as a separate tensor entirely without changing the input tensor's values.

\subsubsection{Function: {\tt corecontractor}}

The tensors must now be converts to their matrix equivalent representation having defined a permutation and reshaping of the tensors. This function {\tt corecontractor} will manipulate both tensors according to their contracted indices and their conjugation.

\begin{lstlisting}[language=julia]
function corecontractor(conjA::Bool,conjB::Bool,A::densTensType,iA::intvecType,
			B::densTensType,iB::intvecType,Z::densTensType...;
			alpha::Number=1.,beta::Number=1.)
  notvecA = findnotcons(ndims(A),iA)
  C = prepareT(A,notvecA,iA,conjA)
  notvecB = findnotcons(ndims(B),iB) 
  D = prepareT(B,iB,notvecB,conjB)

  if length(Z) > 0
    tempZ = reshape(Z[1],size(C,1),size(D,2))
    CD = libmult(C,D,tempZ...,alpha=alpha,beta=beta)
  else
    CD = libmult(C,D,alpha=alpha,beta=beta)
  end

  AAsizes = ()
  for w = 1:length(notvecA)
    AAsizes = (AAsizes...,size(A,notvecA[w]))
  end
  for w = 1:length(notvecB)
    AAsizes = (AAsizes...,size(B,notvecB[w]))
  end
  return CD,AAsizes
end
\end{lstlisting}

The function {\tt findnotcons} was already defined in {\tt tensor.jl} and returns the complement tuple to the input contracted indices.  It represents the uncontracted indices.

\subsubsection{Function: {\tt maincontractor}}

In the previously defined front-end contractors, each led to the {\tt maincontractor} function.  This will receive an argument indicating that it will be conjugated or not ({\tt conjA} and {\tt conjB}).

\begin{lstlisting}[language=julia]
function maincontractor(conjA::Bool,conjB::Bool,A::AbstractArray,
		      iA::intvecType,B::AbstractArray,iB::intvecType,
		      Z::AbstractArray...;alpha::Number=1.,
		      beta::Number=1.)
  CD,AAsizes = corecontractor(conjA,conjB,A,iA,B,iB,Z...,alpha=alpha,beta=beta)
  return reshape(CD,AAsizes)
end

function maincontractor(conjA::Bool,conjB::Bool,A::tens{X},iA::Tuple,
		      B::tens{Y},iB::Tuple,Z::tens{W}...;alpha::Number=1.,
		      beta::Number=1.) where {X <: Number, Y <: Number, 
		      W <: Number}
  CD,AAsizes = corecontractor(conjA,conjB,A,iA,B,iB,Z...,alpha=alpha,beta=beta)
  outType = typeof(X(1)*Y(1))
  nelem = prod(size(CD))
  return tens{outType}(AAsizes,reshape(CD,nelem))
end
export maincontractor
\end{lstlisting}

The reason for calling yet another function {\tt corecontractor} is because beyond this point, the functions for the {\tt AbstractArray} type and {\tt denstens} converge.

At this point, having used so much of julia's machinery to compute the contraction, it might be asked why bother defining the tensor with the {\tt denstens} type.  The answer is that neglecting the rank of the tensor allows for better compilation of the code by julia's compiler, especially when a tensor is being reshaped frequently.  This feature also makes it easier to handle at the higher levels.  Since this function only requires a a matrix input, then this makes a natural point where julia's linear algebra operations become efficient to use. This also carries a message for lower level implementations. The conversion of the tensor to matrices for contraction (and also decomposition) means that programs to compute the matrix multiplication can be used as this point. 

In terms of the DMRG algorithm, the contraction step is often practically the most expensive. The decomposition step will formally scale worse than the contraction step, but the need to compute several contractions (6 in a straightforward implementation of the two-site algorithm \cite{dmrjulia1}) to form the reduced site representation of Schr\"odinger's equation at each step. These tend to be where the code will spend most of its time, so it is important to ensure the contraction algorithm is efficient.

\subsection{Contraction functions}

The contract function is one of the four ore operations in a tensor network library as defined in Ref.~\onlinecite{bakerCJP21}. One overarching point about the implementation of the contract function here is that the code will not automatically catch errors.  One strategy is to use these functions with julia's {\tt Array} type which will find errors in the input to some degree. If julia implements a good method for disabling error messages, then this will be considered in a future version.

\subsubsection{Functions: {\tt contract}, {\tt ccontract}, {\tt contractc}, and {\tt ccontractc} (scalar output)}

There are ways to ensure that the conjugation of a tensor is done efficiently if a tensor and one of the tensors needs to be conjugated.  In order to account for the four types of contractions with and without conjugation between two tensors, the functions {\tt contract}, {\tt ccontract}, {\tt contractc}, and {\tt ccontractc} are defined to be dependent on a {\tt maincontractor} function that will be defined above.

If the result of the contraction has no indices, it is useless to define the input indices.  Inputting only the two tensors to be contracted will automatically assume that the tensors are correctly permuted and then output a scalar value.

\begin{lstlisting}[language=julia]
function contract(A::TensType,B::TensType;alpha::Number=1.)
  mA,mB = checkType(A,B)
  vec_in = ntuple(i->i,ndims(mA))
  C = contract(mA,vec_in,mB,vec_in,alpha=alpha)
  return searchindex(C,1,1)
end

function ccontract(A::TensType,B::TensType;alpha::Number=1.)
  mA,mB = checkType(A,B)
  vec_in = ntuple(i->i,ndims(mA))
  C = ccontract(mA,vec_in,mB,vec_in,alpha=alpha)
  return searchindex(C,1,1)
end

function contractc(A::TensType,B::TensType;alpha::Number=1.)
  mA,mB = checkType(A,B)
  vec_in = ntuple(i->i,ndims(mA))
  C = contractc(mA,vec_in,mB,vec_in,alpha=alpha)
  return searchindex(C,1,1)
end

function ccontractc(A::TensType,B::TensType;alpha::Number=1.)
  mA,mB = checkType(A,B)
  vec_in = ntuple(i->i,ndims(mA))
  C = ccontractc(mA,vec_in,mB,vec_in,alpha=alpha)
  return searchindex(C,1,1)
end
\end{lstlisting}

One additional feature that is often useful for debugging (or to directly access the norm squared of a tensor) is to admit one single tensor as an argument.  This essentially computes $A\cdot A$ (contraction of a tensor $A$ along all indices or conjugating one of the two terms with {\tt ccontract} or {\tt contractc}). This 'self-contraction' function is only useful for coding purposes in some cases, but the use of {\tt norm} and then squaring the output is perfectly acceptable in this case and produces identical results.

\begin{lstlisting}[language=julia]
function contract(A::TensType;alpha::Number=1.)
  return contract(A,A,alpha=alpha)
end

function ccontract(A::TensType;alpha::Number=1.)
  return ccontract(A,A,alpha=alpha)
end

function contractc(A::TensType;alpha::Number=1.)
  return contractc(A,A,alpha=alpha)
end

function ccontractc(A::TensType;alpha::Number=1.)
  return ccontractc(A,A,alpha=alpha)
end
\end{lstlisting}

\subsubsection{Functions: {\tt contract}, {\tt ccontract}, {\tt contractc}, and {\tt ccontractc} (tensor output)}

For each of the functions that result in a tensor (some indices remain), a prefactor can be applied as well as an additional tensor added onto the contraction.  This is to mirror the axpy functions that are commonly in many libraries for matrix multiplication,
\begin{equation}
C = \alpha A\cdot B+\beta Z
\end{equation}
but the multiplications here should be generalized to tensor contractions.

The inputs {\tt iA} and {\tt iB} can be any of the {\tt genvecType} but will ultimately be converted to a tuple of integers. The reason for using a tuple is because it can be more efficient in the view of the compiler, although the difference here is negligible as was discussed for {\tt convIn}.

\begin{lstlisting}[language=julia]
function contract(A::TensType,iA::intvecType,B::TensType,iB::intvecType,
			Z::TensType...;alpha::Number=1.,beta::Number=1.)
  mA,mB = checkType(A,B)
  return maincontractor(false,false,mA,convIn(iA),mB,convIn(iB),Z...,
  			alpha=alpha,beta=beta)
end

function ccontract(A::TensType,iA::intvecType,B::TensType,iB::intvecType,	
			Z::TensType...;alpha::Number=1.,beta::Number=1.)
  mA,mB = checkType(A,B)
  return maincontractor(true,false,mA,convIn(iA),mB,convIn(iB),Z...,
  			alpha=alpha,beta=beta)
end

function contractc(A::TensType,iA::intvecType,B::TensType,iB::intvecType,	
			Z::TensType...;alpha::Number=1.,beta::Number=1.)
  mA,mB = checkType(A,B)
  return maincontractor(false,true,mA,convIn(iA),mB,convIn(iB),Z...,
  			alpha=alpha,beta=beta)
end

function ccontractc(A::TensType,iA::intvecType,B::TensType,iB::intvecType,
			Z::TensType...;alpha::Number=1.,beta::Number=1.)
  mA,mB = checkType(A,B)
  return maincontractor(true,true,mA,convIn(iA),mB,convIn(iB),Z...,
  			alpha=alpha,beta=beta)
end
export contract,ccontract,contractc,ccontractc
\end{lstlisting}

Recall that the use of {\tt checkType} is essentially zero if the same two types are input into the program. It is conceded that this implementation is especially easy in julia where types can be defined generally. For lower level implementations, more function calls may be required. However, for learning purposes, this is sufficient.

Note the use of {\tt maincontractor} as a unifying function here. The boolean inputs will determine where the tensors are conjugated (when constructing the matrix equivalent).

Note that the {\tt Z} tensor input does not need to be present even though it appears in the function call.  This is the same for {\tt alpha} and {\tt beta}. So, all of
\begin{lstlisting}[numbers=none]
contract(A,(1,2),B,(1,2))
contractc(A,(1,2),B,(1,2),alpha=3)
contract(A,[1,2],B,[1,2])
contract(A,[1,2],B,[1,2],Z)
contract(A,[1,2],B,[1,2],Z,alpha=3.,beta=4)
\end{lstlisting}
are valid and more combinations are possible. Note that if two indices are reshaped together on one tensor and that the other tensor has not, then the call
\begin{lstlisting}[numbers=none]
A = rand(20,30,40)
B = rand(20,30,40)
rB = reshape(B,[[1],[2,3]])
C = contract(A,(2,3),rB,(2,)) # = contract(A,(2,3),B,(2,3))
\end{lstlisting}
is valid.

\subsubsection{Functions: {\tt contract}, {\tt ccontract}, {\tt contractc}, and {\tt ccontractc} (re-ordered tensor output)}

One very frequently used feature of the contract function is to reorder the tensor once it has been contracted.  This is accomplished by adding the reordering vector as the first argument, {\tt order}.

\begin{lstlisting}[language=julia]
function contract(order::intvecType,A::TensType,iA::intvecType,B::TensType,
		iB::intvecType,Z::TensType...;alpha::Number=1.,beta::Number=1.)
  mA,mB = checkType(A,B)
  newT = contract(mA,iA,mB,iB,Z...,alpha=alpha,beta=beta)
  return permutedims!(newT,convIn(order))
end

function ccontract(order::intvecType,A::TensType,iA::intvecType,B::TensType,
		iB::intvecType,Z::TensType...;alpha::Number=1.,beta::Number=1.)
  mA,mB = checkType(A,B)
  newT = ccontract(mA,iA,mB,iB,Z...,alpha=alpha,beta=beta)
  return permutedims!(newT,convIn(order))
end

function contractc(order::intvecType,A::TensType,iA::intvecType,B::TensType,
		iB::intvecType,Z::TensType...;alpha::Number=1.,beta::Number=1.)
  mA,mB = checkType(A,B)
  newT = contractc(mA,iA,mB,iB,Z...,alpha=alpha,beta=beta)
  return permutedims!(newT,convIn(order))
end


function ccontractc(order::intvecType,A::TensType,iA::intvecType,B::TensType,
		iB::intvecType,Z::TensType...;alpha::Number=1.,beta::Number=1.)
  mA,mB = checkType(A,B)
  newT = ccontractc(mA,iA,mB,iB,Z...,alpha=alpha,beta=beta)
  return permutedims!(newT,convIn(order))
end
\end{lstlisting}

An example of how to use this function is given here with the equivalent permute after the contraction is evaluated.

\begin{lstlisting}[numbers=none]
A = rand(15,20,5)
B = rand(5,10,20)
C = contract(A,(3,2),B,(1,3))
D = permutedims!(C,[2,1])
#or
D = contract([2,1],A,(3,2),B,(1,3))
\end{lstlisting}

\subsubsection{Function: {\tt trace}}

Where the {\tt contract} function had a self-contract function that contracted over all indices in a tensor with a copy of that tensor, there is also the need to perform a partial trace over some of the indices on a tensor. This will be handled by the trace function.  For example, 
\begin{lstlisting}[numbers=none]
A = rand(10,20,40,30,10,5,20)
B = trace(A,[[1,5],[2,7]])
\end{lstlisting}
will trace over the indices first and fifth indices on {\tt A} as well as the second and seventh indices. For a single set of indices, the command can reduce to 
\begin{lstlisting}[numbers=none]
B = trace(A,[1,5])
\end{lstlisting}
This function makes use of the {\tt makeId} function that was introduced earlier to accomplish the necessary contractions.  Essentially, an identify matrix (or tensor) is contracted onto the indices in question.

\begin{lstlisting}[language=julia]
function trace(A::TensType,iA::W) where W <: Union{intvecType,AbstractArray}
  if typeof(iA) <: intvecType
    Id = makeId(A,iA)
    conA = iA
  else
    Id = makeId(A,iA)
    conL = [iA[w][1] for w = 1:length(iA)]
    conR = [iA[w][2] for w = 1:length(iA)]
    conA = vcat(conL,conR)
  end
  return contract(A,conA,Id,[i for i = 1:ndims(Id)])
end
export trace
\end{lstlisting}

\subsubsection{Function: {\tt checkcontract}}

It is sometimes the case that the contraction must be checked as a part of debugging a program. The {\tt checkcontract} function does this.  Remember that DMRjulia does not implement inherent error checks and instead provides functions such as {\tt checkcontract} to accomplish this.  The dense case simply tests the sizes of the indices that are being contracted over are the same. The quantum number case also implements a check for the fluxes of each index \cite{dmrjulia2}.

\begin{lstlisting}[language=julia]
function checkcontract(A::TensType,iA::intvecType,B::TensType,iB::intvecType,
		Z::TensType...;alpha::Number=1.,beta::Number=1.)
  mA,mB = checkType(A,B)

  if size(mA)[iA] == size(mB)[iB]
    println("contracting over indices with equal sizes")
  else
    error("some indices in A or B are not equal size; A->",
    		size(mA)[iA],", B->",size(mB)[iB])
  end
  if typeof(mA) <: qarray
    println("checking flux:")
    checkflux(mA)
    checkflux(mB)
    for a = 1:length(iA)
      AQNs = recoverQNs(iA[a],A)
      BQNs = recoverQNs(iB[a],B)
      println("contracted index $a (A's index: ",iA[a],
      				", B's index: ",iB[a],")")
      if length(AQNs) == length(BQNs)
        for w = 1:length(AQNs)
          if AQNs[w] != inv(BQNs[w])
            error("unmatching quantum numbers on indices on index ",
		iA[a]," of A and index ",iB[a]," of B: value ",w)
          end
        end
      else
        error("unmatching quantum number vector lengths")
      end
      println("maching quantum numbers on both indices")
    end
  end
  nothing
end
export checkcontract
\end{lstlisting}

If the function finds not errors, then nothing will happen (except for a slowdown due to checking the relevant quantities). However, if there is an error, the program will be interrupted and a partially explanatory message will be displayed.

\section{Module: {\tt decompositions.jl}}

The module containing the decomposition functions 
%
%
will contain all of {\tt SVD}, {\tt eigen}, and {\tt QR} and {\tt LQ} decompositions.

\subsubsection{Function: {\tt libsvd}}

As of the present version, julia occaisionally uses a lazy permute type for transposes, permutes, and other operations. This means the operation is not performed when the function is called. A new data type is created with special instruction to perform the operation when the language determines that it must be done. It was noticed in an earlier version that this action generated a tremendous slow down in the code because it did not handle parallelization well.  To guard against this, the library's SVD function explicitly converts the transpose of $\hat V$ to a generic Array.

\begin{lstlisting}[language=julia]
function libsvd(X::Array{T,2}) where T <: Number
  F = LinearAlgebra.svd(X)
  return F.U,F.S,Array(F.Vt)
end
export libsvd
\end{lstlisting}

In the online version of this library, another safeguard was added to ensure that some undefined flags would not cancel a computation.  These only appear in a few contrived circumstances and can mostly be resolved by implementing some low-level functions.  However, the basic implementation here does not need them, and they can be guarded against by properly defining the core BLAS or other library's function.

\subsubsection{Function: {\tt recursive\_SVD}}

This recursive version of the SVD will compute all singular values with greater precision. The native SVD implementation in many linear algebra solvers will limit precision based on the largest singular value. 

\begin{lstlisting}[language=julia]
function recursive_SVD(AA::Array{W,2},
	tolerance::Float64=1E-4)::Tuple{Matrix{W},Vector{Float64},Matrix{W}} 
	where W <: Number
  U,D,V = safesvd(AA)
  counter = 2
  anchor = 1
  while counter <= size(D,1)
    if D[counter]/D[anchor] < tolerance
      Lpsi = U[:,counter:size(D,1)]
      Rpsi = V[counter:size(D,1),:]
      X = LinearAlgebra.dot(Lpsi,AA,Rpsi)
      Up,Dp,Vp = safesvd(X)
      for a = counter:size(D,1)
        U[:,a] = sum(p->U[:,p]*Up[p-counter+1,a-counter+1],counter:size(D,1))
        V[a,:] = sum(p->V[p,:]*Vp[p-counter+1,a-counter+1],counter:size(D,1))
      end
      D[counter:size(D,1)] = Dp
      anchor = counter
    end
    counter += 1
  end
  newD = Array(LinearAlgebra.Diagonal(D))
  return U,newD,V
end
\end{lstlisting}

This function is rarely useful for algorithms but may be useful for plotting singular values over a wide range of magnitudes.  It can be called through {\tt svd(...,recursive=true)}

\subsubsection{Function: {\tt findnewm}}

This function obtains the number of singular values to keep on truncation.  The input is the $D$ object from the SVD in a vector form (equivalent to a diagonal matrix), the maximal bond dimension {\tt m}, the magnitude of the original input matrix {\tt mag}, the cutoff parameter {\tt cutoff}, the effective vlaue of zero {\tt effZero} (set to machine precision for doubles of $10^{-16}$ in this library), {\tt nozeros} will remove all approximately zero singular values, {\tt power} is the exponent to which the singular values should be raised to, and the boolean {\tt keepdeg} will avoid truncating degenerate singular values if set to true.

\begin{lstlisting}[language=julia]
function findnewm(D::Array{W,1},m::Integer,minm::Integer,mag::Float64,
		cutoff::Real,effZero::Real,nozeros::Bool,power::Number,
		keepdeg::Bool) where {W <: Number}
  sizeD = size(D,1)
  p = sizeD
  if mag == 0.
    sumD = 0.
    @simd for a = 1:size(D,1)
      @inbounds sumD += abs(D[a])^2
    end
  else
    sumD = mag
  end

  truncerr = 0.

  if cutoff > 0.
    modcutoff = sumD*cutoff
    @inbounds truncadd = abs(D[p])^power
    while p > 0 && ((truncerr + truncadd < modcutoff) || (nozeros 
    		&& abs(D[p]) < effZero))
      truncerr += truncadd
      p -= 1
      @inbounds truncadd = abs(D[p])^power
    end
    if keepdeg
      while p < length(D) && isapprox(D[p],D[p+1])
        p += 1
      end
    end
    thism = m == 0 ? max(min(p,sizeD),minm) : max(min(m,p,sizeD),minm)
  else
    thism = m == 0 ? max(sizeD,minm) : max(min(m,sizeD),minm)
  end

  return thism,sizeD,truncerr,sumD
end
\end{lstlisting}

This function also applies to the truncation of eigenvalues, often where {\tt power = 1} in that case.

\subsubsection{Function: {\tt safesvd}}

For all practical purposes, this function is equivalent to {\tt LinearAlgebra.svd}.  Some incompatibilities for highly degenerate input matrices were noticed, so backup implementations of the SVD from Ref.~\onlinecite{dmrjulia1} were implemented to solve those issues.  At the present time, julia is not fully interfaced with the BLAS library's implementation of the SVD in that some options are pre-selected.  In principle, choosing these options fully could make the SVD work here, but for safety and backward compatbility, this function will remain here.  There has been no testable consequence for this by us and the library's implementation works on all tested cases.

See the current code version for the current implementation of this function \cite{dmrjulia}.


%
%

\subsubsection{Function: {\tt svd}}\label{svdfunction}

The primary SVD function truncates the singular values according to the discussion in the text.  The power that the singular values are taken to before truncation can be adjusted, but there is rarely a use to do this.  Instead, the function is uniform between the SVD and the eigenvalue solver below, which justifies its purpose.

\begin{lstlisting}[language=julia]
function svd(AA::Array{T,2};power::Number=2,cutoff::Float64 = 0.,
		m::Integer = 0,mag::Float64=0.,minm::Integer=2,
		nozeros::Bool=false,recursive::Bool = false,
		effZero::Float64 = 1E-16,keepdeg::Bool=false) where T <: Number
    U,D,Vt = recursive ? recursive_SVD(AA) : safesvd(AA)
    thism,sizeD,truncerr,sumD = findnewm(D,m,minm,mag,cutoff,effZero,nozeros,
    					power,keepdeg)
    if sizeD > thism
      @inbounds Utrunc = U[:,1:thism]
      @inbounds Dtrunc = D[1:thism]
      @inbounds Vtrunc = Vt[1:thism,:]
    elseif sizeD < thism
      Utrunc = zeros(eltype(U),size(U,1),thism)
      Dtrunc = zeros(eltype(D),thism)
      Vtrunc = zeros(eltype(Vt),thism,size(Vt,2))
      @inbounds Utrunc[:,1:size(U,2)] = U
      @inbounds Dtrunc[1:size(D,1)] = D
      @inbounds Vtrunc[1:size(Vt,1),:] = Vt
    else
      Utrunc = U
      Dtrunc = D
      Vtrunc = Vt
    end
    Darray = Array(LinearAlgebra.Diagonal(Dtrunc))::Array{Float64,2}
    return Utrunc,Darray,Vtrunc,truncerr,sumD
end
export svd
\end{lstlisting}

There are many aspects of this function that are discussed here. For the physics interpretation of the decomposition or some of the approximation parameters such as {\tt cutoff}, the reader is directed to Ref.~\onlinecite{dmrjulia1}.

The options for this function can be useful in many situations. The parameter {\tt power} simply gives the effective norm at which the singular values should be truncated (here, the values are squared since those are the values of the density matrix \cite{dmrjulia1}). The parameter {\tt cutoff} is the maximum sum of singular values to truncate that can override the {\tt m} parameter (maximum bond dimension of the newly introduced index).  

The {\tt mag} is the Froebenius norm of the input tensor and is the value that the cutoff is scaled to. This often matters when truncating two tensors that are joined together.  There is a miniscule cost savings with setting this parameter, but it can be done and avoid the computation (for example, the two-site DMRG algorithm is always norm 1 for all tensors, so this can be set to be unity always).

Mainly for testing purposes, the {\tt minm} parameter can be set to guarantee a bond dimension greater than {\tt minm}. This can be have a use in other contexts, but it is rare to need it.  The {\tt dmrg} function does set {\tt minm=2} by default because it is assumed that all models will search for an entangled ground state which guarantees $m\geq2$. This is often useful for stability reasons but again is not strictly necessary. There is no problem using this with a purely classical model with bond dimension of 1 everywhere.

There are also sometimes when only the non-zero values of the SVD should be kept and this can be toggled with {\tt nozeros}. If the recursive SVD from Ref.~\onlinecite{stoudenmire2013real} is demanded, then {\tt recursive} can be toggled to true. The recursive SVD is almost never required for an algorithm, although it can be useful for displaying highly accurate singular values.  

The {\tt effZero} parameter specifies the effective zero value for the computation and {\tt keepdeg} will not truncate degenerate singular values.

Note that the output arrays are ensured to be of the {\tt Array} type.  A large slowdown was noticed when using the special array types in julia. Conversion to the array types did not create a large slowdown or excessive allocations, so this was used instead of possibly causing a single-thread slow-down for some large computation later on.

The outputs from the {\tt svd} function are the $U$, $D$, and $V^\dagger$ tensors with some additional information that can be useful. The fourth output is the truncation error. The final value is the magnitude.  The julia language allows for the specifying of only (for example) three of the outputs and will leave off those not specified. If one output is set, then a tuple is stored in that variable with all five outputs, so the previous trick of reducing the outputs will work for 2 or more outputs here.

There is a similar function defined for the {\tt denstens} input

\begin{lstlisting}[language=julia]
function svd(AA::tens{W};power::Number=2,cutoff::Float64 = 0.,
          m::Integer = 0,mag::Float64=0.,minm::Integer=2,nozeros::Bool=false,
          recursive::Bool = false,effZero::Number = 1E-16,
          keepdeg::Bool=false) where W <: Number
  rAA = reshape(AA.T,size(AA)...)
  U,D,V,truncerr,sumD = svd(rAA,power=power,cutoff=cutoff,m=m,mag=mag,
  			minm=minm,nozeros=nozeros,recursive=recursive,
			effZero=effZero,keepdeg=keepdeg)
  tensU = tens(W,U)
  tensD = tens(Float64,D)
  tensV = tens(W,V)
  return tensU,tensD,tensV,truncerr,sumD
end
\end{lstlisting}

One convenient format for the SVD is to group indices as was done for the {\tt reshape} function. An input such as {\tt svd(A,[[1,2],[3]])} will group together indices 1 and 2, perform the SVD, and then unreshape those indices on the returning $\hat U$ tensor.

\begin{lstlisting}[language=julia]
function svd(AA::TensType;power::Number=2,cutoff::Float64 = 0.,
    	m::intType = 0,mag::Float64=0.,minm::intType=2,nozeros::Bool=false,
    	recursive::Bool = false) where T <: Number
  AB,Lsizes,Rsizes = getorder(AA,vecA)
  U,D,V,truncerr,newmag = svd(AB,power = power,cutoff=cutoff,m=m,mag=mag,
  				minm=minm,nozeros=nozeros,recursive=recursive,
				keepdeg=keepdeg)

  outU = unreshape!(U,Lsizes...,size(D,1))
  outV = unreshape!(V,size(D,2),Rsizes...)
  return outU,D,outV,truncerr,newmag
end
\end{lstlisting}

\subsubsection{Function: {\tt libeigen}}

The eigenvalue solver used in the library assumes a symmetric matrix input.  This is both for efficiency in julia and because it is the expected case. The function handles both orthogonal basis sets (regular eigenvalue decomposition) and problems requiring a generalized eigenvalue problem.

\begin{lstlisting}[language=julia]
function libeigen(AA::Array{W,2},B::Array{W,2}...) where W <: Number
  if eltype(AA) <: Complex
    M = LinearAlgebra.Hermitian((AA+AA')/2)
  else
    M = LinearAlgebra.Symmetric((AA+AA')/2)
  end
  return length(B) == 0 ? LinearAlgebra.eigen(M) : LinearAlgebra.eigen(M,B[1])
end
\end{lstlisting}

Note that this implementation will therefore assumes Hermitian input Hamiltonians.

\subsubsection{Function: {\tt eigen}}

The eigenvalue decomposition can be similarly defined to the SVD. The truncation on the $\hat D$ matrix is done without squaring the value as was done for the SVD.

\begin{lstlisting}[language=julia]
function eigen(AA::Array{W,2},B::Array{W,2}...;cutoff::Float64 = 0.,
		m::Integer = 0,mag::Float64=0.,minm::Integer=1,
		nozeros::Bool=false,power::Number=1,
		effZero::Float64 = 1E-16,
		keepdeg::Bool=false) where {W <: Number}

  Dsq,U = libeigen(AA,B...)

  thism,sizeD,truncerr,sumD = findnewm(Dsq,m,minm,mag,cutoff,effZero,
  					nozeros,power,keepdeg)
  if sizeD > thism
    @inbounds Utrunc = U[:,sizeD-thism+1:sizeD]
    @inbounds Dtrunc = Dsq[sizeD-thism+1:sizeD]
  elseif sizeD < minm
    Utrunc = zeros(eltype(U),size(U,1),minm)
    Dtrunc = zeros(eltype(Dsq),minm)
    @inbounds Utrunc[:,1:thism] = U
    @inbounds Dtrunc[1:thism] = Dsq
  else
    Utrunc = U
    Dtrunc = Dsq
  end
  finalD = Array(LinearAlgebra.Diagonal(Dtrunc))::Array{Float64,2}
  return finalD,Utrunc,truncerr,sumD
end
export eigen
\end{lstlisting}

Note that the eigenvalue decomposition is taken to be symmetric in the center.

The inputs are nealry identical to the {\tt svd} function. The exception is that the overlap matrix {\tt B} can be provided in the even that a generalized eigenvalue problem is to be solved of the form
\begin{equation}
\mathcal{H}\psi=E\hat B\psi
\end{equation}

The function is also defined for a {\tt denstens} input

\begin{lstlisting}[language=julia]
function eigen(AA::tens{W},B::tens{R}...;cutoff::Float64 = 0.,
              m::Integer = 0,mag::Float64=0.,minm::Integer=2,
              nozeros::Bool=false,recursive::Bool = false,
              keepdeg::Bool=false) where {W <: Number,R <: Number}
  X = reshape(AA.T,size(AA)...)
  D,U,truncerr,sumD = eigen(X,B...,cutoff=cutoff,m=m,mag=mag,minm=minm,
  				nozeros=nozeros,keepdeg=keepdeg)
  tensD = tens(Float64,D)
  tensU = tens(T,U)
  return tensD,tensU,truncerr,mag
end
\end{lstlisting}

Similarly for the {\tt reshape} and {\tt svd} functions, the eigenvalue decomposition can conveniently group indices together and then unreshape them. An input such as {\tt eigen(A,[[1,2],[3]])} will group together indices 1 and 2, perform the eigenvalue decomposition, and then unreshape those indices on the returning $\hat U$ tensor.

\begin{lstlisting}[language=julia]
function eigen(AA::TensType,vecA::Array{Array{W,1},1},
                B::TensType...;cutoff::Float64 = 0.,m::Integer = 0,
                mag::Float64=0.,minm::Integer=1,nozeros::Bool=false,
                keepdeg::Bool=false) where W <: Integer
  AB,Lsizes,Rsizes = getorder(AA,vecA)
  D,U,truncerr,newmag = eigen(AB,B...,cutoff=cutoff,m=m,mag=mag,minm=minm,
  				nozeros=nozeros,keepdeg=keepdeg)
  outU = unreshape!(U,Lsizes...,size(D,1))
  return D,outU,truncerr,newmag
end 
\end{lstlisting}

\subsubsection{Function: {\tt libqr} \& {\tt liblq}}

The QR decomposition and LQ decompositions are implemented here using the {\tt LinearAlgebra} package.

\begin{lstlisting}[language=julia]
function libqr(R::TensType;decomposer::Function=LinearAlgebra.qr)
  A = makeArray(R)
  P,Q = decomposer(A)
  if size(P,2) > size(Q,1)
    U = P[:,1:size(Q,1)]
    V = Array(Q)
  elseif size(P,2) < size(Q,1)
    U = Array(P)
    V = Q[1:size(P,2),:]
  else
    U = Array(P)
    V = Array(Q)
  end
  return U,V
end

function liblq(R::TensType;decomposer::Function=LinearAlgebra.lq)
  return libqr(R,decomposer=decomposer)
end
\end{lstlisting}

The LQ decomposition can be defined very nearly the same way but by changing out the function from julia's {\tt LinearAlgebra} package.  It turns out that in this version of julia, using a function as an argument has little overhead.

\subsubsection{Function: {\tt qr} \& {\tt lq}}

The LQ and QR decompositions produce a unitary tensor $\hat Q$ and another weighted tensor $\hat L$ or $\hat R$. One realization of the LQ and QR decompositions is the SVD with the $\hat D$ contracted onto one of the $\hat U$ or $\hat V^\dagger$ tensors.  However, the LQ and QR decomposition have the special property that $\hat R$ ($\hat L$) is upper (lower) triangular.

It is in principle possible to truncate the decomposition, but this will not be implemented here. The return values for the truncation error and magnitude will be set to 0 and 1 respectively.

\begin{lstlisting}[language=julia]
function qr(AA::TensType,vecA::Array{Array{W,1},1}) where W <: Integer
  AB,Lsizes,Rsizes = getorder(AA,vecA)
  Qmat,Rmat,truncerr,newmag = qr(AB)

  innerdim = size(Qmat,2)

  outU = unreshape!(Qmat,Lsizes...,innerdim)
  outV = unreshape!(Rmat,innerdim,Rsizes...)
  return outU,outV,truncerr,newmag
end

function qr(AA::AbstractArray;decomposer::Function=LinearAlgebra.qr)
  return libqr(AA,decomposer=decomposer)
end

function qr(AA::denstens;decomposer::Function=LinearAlgebra.qr)
  rAA = makeArray(AA)
  Q,R = libqr(rAA,decomposer=decomposer)
  return tens(Q),tens(R),0.,1.
end

function lq(AA::TensType,vecA::Array{Array{W,1},1}) where W <: Integer
  AB,Lsizes,Rsizes = getorder(AA,vecA)
  Lmat,Qmat,truncerr,newmag = lq(AB)

  innerdim = size(Qmat,1)

  outU = unreshape!(Lmat,Lsizes...,innerdim)
  outV = unreshape!(Qmat,innerdim,Rsizes...)
  return outU,outV,truncerr,newmag
end

function lq(AA::AbstractArray;decomposer::Function=LinearAlgebra.lq)
  return libqr(AA,decomposer=decomposer)
end

function lq(AA::denstens;decomposer::Function=LinearAlgebra.lq)
  rAA = makeArray(AA)
  L,Q = libqr(rAA,decomposer=decomposer)
  return tens(L),tens(Q),0.,1.
end
\end{lstlisting}

Passing functions as arguments has very little overhead in julia, hence the definition of {\tt lq} being based on {\tt qr}.  Just as with the SVD, a vector of indices to group together can be input to these functions.

\subsubsection{Function: {\tt polar}}

The polar decomposition can be thought of as a post-processing step onto the SVD.  The purpose is to maintain the outer basis states on the inner index generated by the SVD.  There are two types of decompositions that can be obtained,
\begin{equation}
\hat U\hat D\hat U^\dagger,\hat U\hat V^\dagger\quad\mathrm{or}\quad\hat UV^\dagger,\hat V\hat D\hat V^\dagger
\end{equation}
which can be selected by specifying {\tt right = false} (left form) or {\tt right = true} (right form).

The {\tt group} of indices must be specified similar to the vector input formats for both the {\tt svd} and {\tt eigen} functions. The last new option is to specify the {\tt outermaxm} which allows for a truncation on the outer index.  This is rarely used since the tensor should be preserved but it can limit the inner bond dimension.  All other options are the same as the SVD functions defined earlier.

\begin{lstlisting}[language=julia]
function polar(AA::TensType,group::Array{Array{W,1},1};
                right::Bool=true,cutoff::Float64 = 0.,m::Integer = 0,
                mag::Float64 = 0.,minm::Integer=1,nozeros::Bool=false,
                recursive::Bool=false,outermaxm::Integer=0,
                keepdeg::Bool=false) where W <: Integer
  AB,Lsizes,Rsizes = getorder(AA,group)
  U,D,V,truncerr,newmag = svd(AB,cutoff=cutoff,m=m,mag=mag,minm=minm,
  				nozeros=nozeros,keepdeg=keepdeg)
  U = unreshape!(U,Lsizes...,size(D,1))
  V = unreshape!(V,size(D,2),Rsizes...)
  if right
    if outermaxm > 0
      outermtrunc = m>0 ? min(m,outermaxm,size(V,2)) : min(outermaxm,size(V,2))
      truncV = getindex(V,:,1:outermtrunc)
      modV = V
    else
      truncV = V
      modV = V
    end
    DV = contract(D,2,truncV,1)
    rR = ccontract(modV,1,DV,1)
    rU = contract(U,3,modV,1)
    leftTensor,rightTensor = rU,rR
  else
    if outermaxm > 0
      outmtrunc = m>0 ? min(m,size(U,1),outermaxm) : min(outermaxm,size(U,1))
      truncU = getindex(U,1:outmtrunc,:)
      modU = U
    else
      truncU = U
      modU = U
    end   
    UD = contract(truncU,2,D,1)
    lR = contractc(UD,2,U,2)
    lU = contract(U,2,V,1)
    leftTensor,rightTensor = lR,lU
  end
  return leftTensor,rightTensor,D,truncerr
end
export polar
\end{lstlisting}

%
%


%
%

\section{Module: {\tt MPtask.jl}}

This module will initialize the MPS and MPO for a given system. There is an interest in preserving the memory footprint of the system on a computer in terms of random access memory (RAM).  To reduce the size of the tensors stored, the tensors of the MPS and MPO can be written to a disk and only called when necessary.  Remarkably, this strategy works well even with an expected increase in garbage collection time (most likely due to a forced garbage collection step when the tensor is written or read).  Further remarkably, the overhead costs very little time when used.  Even further remarkably, no high level functions need to be rewritten for this feature to appear, so just as will be seen for quantum number tensors, this improvement adds almost no practical hurdle to implementation, but it will be required for some of the larger systems.



\subsubsection{Types: {\tt MPS}, {\tt MPO}, and {\tt Env}}

There are several types that should be defined for defining the MPS, MPO, and the environment tensors.  It is not strictly required to define the environment tensors as a new type.  One could use the basic definition in julia, yet defining this type is much easier to think through in terms of the code and it makes the code compatible with the large types defined below.

For now, all tensors will be kept in the available memory (not the hard disk) and be regarded as ``regular" (hence the prefix {\tt reg} found in some places).  The types to define are {\tt MPS}, {\tt MPO}, and {\tt Env} (referring to the environment.

\begin{lstlisting}[language=julia]
abstract type MPS end
export MPS


abstract type MPO end
export MPO


abstract type regMPS <: MPS end
export regMPS


abstract type regMPO <: MPO end
export regMPO


abstract type envType end
export envType

abstract type regEnv <: envType end
export regEnv

const Env = Union{AbstractArray,envType}
export Env
\end{lstlisting}

\subsubsection{Struct: {\tt matrixproductstate}}

The MPS is contained in a series of tensors that have two fields.  One is the tensor {\tt A} (named after the traditional pen and paper notation) and the orthogonality center {\tt oc}.

\begin{lstlisting}[language=julia]
mutable struct matrixproductstate{W} <: regMPS where W <: Array{TensType,1}
  A::W
  oc::intType
end
\end{lstlisting}

Whenever needed, this struct's name is never written out explicitly.  Instead, the generic type {\tt MPS} is used.

\subsubsection{Struct: {\tt matrixproductoperator}}

Similar to the {\tt matrixproductstate}, the MPO is contained in the {\tt matrixproductoperator} struct.  There is no guage condition on the MPO, so it is not a field that is stored in the struct.

\begin{lstlisting}[language=julia]
mutable struct matrixproductoperator{W} <: regMPO where W <: Array{TensType,1}
  H::W
end
\end{lstlisting}

Whenever needed, this struct's name is never written out explicitly.  Instead, the generic type {\tt MPO} is used.

\subsubsection{Struct: {\tt environment} \& {\tt vecenvironment}}

The {\tt environment} could have been defined synonymously with the {\tt matrixproductoperator}, but it is useful to have a separate name for this particular field. It is therefore included here as the following.  There is no established symbol for the environment tensor, so the {\tt V} is borrowed for the most pronounced letter of the word.

\begin{lstlisting}[language=julia]
mutable struct environment{W} <: regEnv where W <: Array{TensType,1}
  V::Array{W,1}
end

mutable struct vecenvironment{W} <: regEnv where W <: Array{TensType,1} 
  V::Array{Array{W,1}}
end
export vecenvironment
\end{lstlisting}

Whenever needed, this struct's name is never written out explicitly.  Instead, the generic type {\tt Env} is used.  The main reason for defining the vector of environments ({\tt vecenvironment}) is to be able to use the global type {\tt Env} for this quantity as well.  There are a few situations where this is useful for general MPS optimization.

\subsubsection{Constructor: {\tt environment}}

To make an environment from a list of tensors, the following function can be defined.

\begin{lstlisting}[language=julia]
function environment(T::TensType...)
  return environment([T...])
end
export environment
\end{lstlisting}

\subsubsection{Constructor: {\tt makeoc}}

The orthogonality center must be defined on the lattice.  In one of the rare instances where a check is implemented, the {\tt makeoc} function will check that the orthogonality center is properly defined and return 1 if nothing is defined.

\begin{lstlisting}[language=julia]
function makeoc(Ns::Integer,oc::intType...)
  return length(oc) > 0  && 0 < oc[1] <= Ns ? oc[1] : 1
end
\end{lstlisting}

\subsubsection{Constructor: {\tt MPS}}\label{MPScomplex}

This constructor will make the MPS with the orthogonality center optionally defined (default: site 1).  The optional {\tt regtens} will produce a julia defined {\tt Array} for the output in case this is necessary.  

\begin{lstlisting}[language=julia]
function MPS(psi::Array{W,1};regtens::Bool=false,oc::Integer=1,
		type::DataType=eltype(psi[1])) where W <: densTensType
  if eltype(psi[1]) != type && !regtens
    MPSvec = [convertTens(type, copy(B[i])) for i = 1:size(B,1)]
  else
    MPSvec = psi
  end
  return matrixproductstate(MPSvec,oc)
end

function MPS(psi::MPS;regtens::Bool=false,oc::Integer=psi.oc,
		type::DataType=eltype(psi))
  return MPS(psi.A,regtens=regtens,oc=oc,type=type)
end

function MPS(B::Array{W,1};regtens::Bool=false,oc::Integer=1,
		type::DataType=eltype(B[1])) where W <: AbstractArray
  if !regtens
    MPSvec = [tens(convert(Array{type,ndims(B[i])},copy(B[i]))) 
    				for i = 1:size(B,1)]
  else
    MPSvec = [convert(Array{type,ndims(B[i])},copy(B[i])) for i = 1:size(B,1)]
  end
  return matrixproductstate{typeof(MPSvec)}(MPSvec,oc)
end

function MPS(type::DataType,B::Union{MPS,Array{W,1}};regtens::Bool=false,
		oc::Integer=1) where W <: AbstractArray
  return MPS(B,regtens=regtens,oc=oc,type=type)
end

function MPS(physindvec::Array{W,1};regtens::Bool=false,oc::Integer=1,
		type::DataType=Float64) where W <: Integer
  Ns = length(physindvec)
  if regtens
    vec = Array{Array{type,3},1}(undef,Ns)
    for w = 1:Ns
      vec[w] = zeros(type,1,physindvec[w],1)
      vec[w][1,1,1] = 1
    end
  else
    vec = Array{tens{type},1}(undef,Ns)
    for w = 1:Ns
      temp = zeros(type,1,physindvec[w],1)
      temp[1,1,1] = 1
      vec[w] = tens(temp)
    end
  end
  return MPS(vec,oc=oc)
end

function MPS(physindvec::Array{W,1},Ns::Integer;regtens::Bool=false,
		oc::Integer=1,type::DataType=Float64) where W <: Integer
  physindvecfull = physindvec[(w-1) % length(physindvec) + 1 for w = 1:Ns]
  return MPS(physindvecfull,regtens=regtens,oc=oc,type=type)
end

function MPS(physindsize::Integer,Ns::Integer;regtens::Bool=false,
		oc::Integer=1,type::DataType=Float64)
  return MPS([physindsize for w = 1:Ns],oc=oc,type=type)
end

function MPS(type::DataType,physindvec::Array{W,1},Ns::Integer;
		regtens::Bool=false,oc::Integer=1) where W <: Integer
  return MPS(physindvec,Ns,regtens=regtens,oc=oc,type=type)
end
\end{lstlisting}

The MPS is often defined to be rank 3 even on the edge tensors.  This form of the MPS constructor could allow for rank-2 on the edges (and then some modification to the eventual DMRG function), but the compiler in julia has an easier time if all the tensors are defined uniformly.  This should be the case in several other languages as well, so it should be considered to keep the rank-3 on each tensor.

Note that a {\tt DataType} can be provided in the first argument which will automatically cause the input arrays to be converted.  This is useful when using complex numbers as the MPS initially defined here is useful for ensuring consistent operations throughout a computation.

Note that the initialization call of the form
\begin{lstlisting}[numbers=none]
physindsize = 2
Ns = 10
psi = MPS(physindsize,Ns)
\end{lstlisting}
generates a matrix product state with physical index size of 2 on each tensor with 10 sites. A vector can also be placed into the first argument with variable physical index sizes.  The state can be initialized with {\tt applyOps}.

\subsection{Function: {\tt applyOps} \& {\tt applyOps!}}

The {\tt applyOps} function provides a quick interface for applying single site operators to the MPS. This can be especially useful for creating starting wavefunctions from the {\tt MPS} initialization of the  starting MPS.

\begin{lstlisting}[language=julia]
function applyOps!(psi::MPS,sites::Array{W,1},Op::TensType;
		trail::TensType=ones(1,1)) where W <: Integer
  for i = 1:length(sites)
    site = sites[i]
    p = site
    psi[p] = contract([2,1,3],Op,2,psi[p],2)
    if trail != ones(1,1)
      for j = 1:p-1
        psi[j] = contract([2,1,3],trail,2,psi[j],2)
      end
    end
  end
  return psi
end
export applyOps!

function applyOps(psi::MPS,sites::Array{W,1},Op::TensType;
		trail::TensType=ones(1,1)) where W <: Integer
  cpsi = copy(psi)
  return applyOps!(cpsi,sites,Op,trail=trail)
end
export applyOps
\end{lstlisting}

\subsubsection{Constructor: {\tt MPO}}

This constructor will make the MPO).  The optional {\tt regtens} will produce a julia defined {\tt Array} for the output in case this is necessary.  

\begin{lstlisting}[language=julia]
function MPO(H::Array{W,1};regtens::Bool=false) where W <: TensType
  T = prod(a->eltype(H[a])(1),1:size(H,1))
  if !regtens && (typeof(H[1]) <: AbstractArray)
    M = [tens(H[a]) for a = 1:length(H)]
  else
    M = H
  end
  return MPO(typeof(T),M,regtens=regtens)
end

function MPO(T::DataType,H::Array{W,1};
		regtens::Bool=false) where W <: TensType
  newH = Array{W,1}(undef,size(H,1))

  for a = 1:size(H,1)
    if ndims(H[a]) == 2
      rP = reshape(H[a],size(H[a],1),size(H[a],2),1,1)
      newH[a] = permutedims!(rP,[4,1,2,3])
    else
      newH[a] = H[a]
    end
  end

  if !regtens && (typeof(newH[1]) <: AbstractArray)
    finalH = [tens(newH[a]) for a = 1:length(newH)]
  else
    finalH = newH
  end
  return matrixproductoperator(finalH)
end

function MPO(T::DataType,mpo::MPO;regtens::Bool=false)
  return MPO(T,mpo.H,regtens=regtens)
end

function MPO(mpo::MPO;regtens::Bool=false)
  return MPO(mpo.H,regtens=regtens)
end
\end{lstlisting}

The MPO can sometimes be useful to define with only matrices (operators) as the constituent tensors.  These will be rank-2 on each site and the constructor which reshapes those tensors formats them into the rank-4 tensor (with trivial horizontal indices of size 1) to make the indices match. 

Just as with the MPS, the MPO can admit a leading {\tt DataType} that converts another MPO or the tensors themselves to another format.

\subsection{Function: {\tt randMPS}}

The function {\tt randMPS} creates a set of tensors with bond dimension 1 on the link indices and randomly chooses which element of the tensor to be assigned a value of 1 (the rest are zero). 

\begin{lstlisting}[language=julia]
function randMPS(T::DataType,physindsize::Integer,Ns::Integer;oc::Integer=1,
		m::Integer=1)
  physindvec = [physindsize for i = 1:Ns]
  return randMPS(T,physindvec,oc=oc,m=m)
end

function randMPS(T::DataType,physindvec::Array{W,1};oc::Integer=1,
		m::Integer=1) where W <: Integer
  Ns = length(physindvec)
  vec = Array{Array{T,3},1}(undef,Ns)
  if m == 1
    for w = 1:Ns
      vec[w] = zeros(1,physindvec[w],1)
      state = rand(1:physindvec[w],1)[1]
      vec[w][1,state,1] = 1
    end
    psi = MPS(vec,oc)
  else
    Lsize,Rsize = 1,prod(w->physindvec[w],2:length(physindvec))
    accumsize = physindvec[1]
    for w = 1:Ns
      physindsize = physindvec[w]
      currRsize = Rsize < accumsize ? min(Rsize,m) : min(accumsize,m)
      vec[w] = rand(T,Lsize,physindsize,currRsize)
      vec[w] /= norm(vec[w])
      Lsize = currRsize
      Rsize = cld(Rsize,physindsize)
      accumsize *= physindsize
    end
    psi = MPS(vec,oc)
    psi[oc] /= expect(psi)
  end
  return psi
end

function randMPS(physindsize::Integer,Ns::Integer;oc::Integer=1,m::Integer=1)
  return randMPS(Float64,physindsize,Ns,oc=oc,m=m)
end

function randMPS(physindvec::Array{W,1};oc::Integer=1,
		m::Integer=1) where W <: Integer
  return randMPS(Float64,physindvec,oc=oc,m=m)
end
export randMPS
\end{lstlisting}

\subsubsection{Elementary functions: {\tt elnumtype}, {\tt eltype}, {\tt size}, {\tt length}, {\tt lastindex}, {\tt copy}, {\tt conj}, {\tt conj!} and {\tt setindex!}}

The elementary functions that were defined for the tensor itself are now defined for the holder types here.

\begin{lstlisting}[language=julia]
function elnumtype(op...)
  opnum = eltype(op[1])(1)
  for b = 2:length(op)
    opnum *= eltype(op[b])(1)
  end
  return typeof(opnum)
end

function size(H::MPO)
  return size(H.H)
end

function size(H::MPO,i::Integer)
  return size(H.H,i)
end

function size(psi::MPS)
  return size(psi.A)
end

function size(psi::MPS,i::Integer)
  return size(psi.A,i)
end

function size(G::regEnv)
  return size(G.V)
end

function size(G::regEnv,i::Integer)
  return size(G.V,i)
end

function length(H::MPO)
  return length(H.H)
end

function length(psi::MPS)
  return length(psi.A)
end

function length(G::regEnv)
  return length(G.V)
end

function eltype(Y::regMPS)
  return eltype(Y.A[1])
end

function eltype(H::regMPO)
  return eltype(H.H[1])
end

function eltype(G::regEnv)
  return eltype(G.V[1])
end

function getindex(A::regMPS,i::Integer)
  return A.A[i]
end

function getindex(A::regMPS,r::UnitRange{W}) where W <: Integer
  if A.oc in r
    newoc = findfirst(w->w == A.oc,r)
  else
    newoc = 0
  end
  return MPS(A.A[r],oc=newoc)
end

function getindex(H::regMPO,i::Integer)
  return H.H[i]
end

function getindex(H::regMPO,r::UnitRange{W}) where W <: Integer
  return MPO(H.H[r])
end

function getindex(G::regEnv,i::Integer)
  return G.V[i]
end

function getindex(G::regEnv,r::UnitRange{W}) where W <: Integer
  return environment(G.V[r])
end

function lastindex(A::regMPS)
  return lastindex(A.A)
end

function lastindex(H::regMPO)
  return lastindex(H.H)
end

function lastindex(G::regEnv)
  return lastindex(G.V)
end

function setindex!(H::regMPO,A::TensType,i::intType)
  H.H[i] = A
  nothing
end

function setindex!(H::regMPS,A::TensType,i::intType)
  H.A[i] = A
  nothing
end

function setindex!(G::regEnv,A::TensType,i::intType)
  G.V[i] = A
  nothing
end

function copy(mps::matrixproductstate{W}) where W <: TensType
  return matrixproductstate{W}([copy(mps.A[i]) for i = 1:length(mps)],
  				oc=copy(mps.oc))
end

function copy(mpo::matrixproductoperator{W}) where W <: TensType
  return matrixproductoperator{W}([copy(mpo.H[i]) for i = 1:length(mpo)])
end

function copy(mps::regMPS)
  return MPS([copy(mps.A[i]) for i = 1:length(mps)],
  				oc=copy(mps.oc))
end

function copy(mpo::regMPO)
  return MPO([copy(mpo.H[i]) for i = 1:length(mpo)])
end

function copy(G::regEnv)
  T = eltype(G[1])
  return envVec{T}([copy(G.V[i]) for i = 1:length(G)])
end

function conj!(A::regMPS)
  conj!.(A.A)
  return A
end

function conj(A::regMPS)
  B = copy(A)
  conj!.(B.A)
  return B
end
\end{lstlisting}

The {\tt length} function returns the number of sites for a given type.  Note also that {\tt elnumtype} returns the type of number stored in a tensor of the MPS, MPO, or environment.  This can be useful in some situations and extends the definition in {\tt tensor.jl}.

\subsubsection{Types: {\tt largeMPS}, {\tt largeMPO}, and {\tt largeEnv}}

Storing large matrices in memory, purely, will cause a difficulty in that only so much memory is available on a computer.  For a truly large tensor, writing it to the disk is useful.  Fortunately, julia has an internal writing protocol to write tensors on the disk. The package, native to the basic functions in julia, is called {\tt Serialization} and makes this process easy.

The abstract types that must be defined are extensions of the previous {\tt reg} types defined above.  Each of {\tt largeMPS}, {\tt largeMPO}, and {\tt largeEnv}.

\begin{lstlisting}[language=julia]
abstract type largeMPS <: MPS end
export largeMPS

abstract type largeMPO <: MPO end
export largeMPO

abstract type largeEnv <: envType end
export largeEnv
\end{lstlisting}

When writing the data to the disk, it is often useful to have an extension.  In this case, ``.dmrjulia" has been chosen.

\begin{lstlisting}[language=julia]
const file_extension = ".dmrjulia"
\end{lstlisting}

\subsubsection{Structs: {\tt largematrixproductstate}, {\tt largematrixproductoperator}, and {\tt largeenvironment}}

Each of {\tt largematrixproductstate}, {\tt largematrixproductoperator}, and {\tt largeenvironment} will store a string denoting a file name and a data type that records what type of data is stored in the tensor.  This second piece of information is only to avoid pulling the tensor from memory when the type is requested.

The large structs are made with the following constructors.  Note the use of the {\tt tensor2disc} to write the tensors.  All three constructors follow the same basic pattern: 1) tensor is written to disk, 2) type of saved for later, 3) file names are saved.  The environment constructor requires both the left and right environments to be made.

\begin{lstlisting}[language=julia]
mutable struct largematrixproductstate <: largeMPS
  A::Array{String,1}
  oc::intType
  type::DataType
end
  
mutable struct largematrixproductoperator <: largeMPO
  H::Array{String,1}
  type::DataType
end

mutable struct largeenvironment <: largeEnv
  V::Array{String,1}
  type::DataType
end
\end{lstlisting}

The MPS also stores the orthogonality center just as the regular version did.

\subsubsection{Structs: {\tt tensor2disc} \& {\tt tensorfromdisc}}

Writing tensors to and from the disk is accomplished here using the {\tt Serialization} package that is internal to julia itself.

\begin{lstlisting}[language=julia]
import Serialization
function tensor2disc(name::String,tensor::TensType;ext::String=file_extension)
  Serialization.serialize(name*ext,tensor)
  nothing
end

function tensorfromdisc(name::String;ext::String=file_extension)
  return Serialization.deserialize(name*ext)
end
\end{lstlisting}

When using this function, a set of files with extension ``.dmrjulia" (default) will appear.

%
%

%
%

\subsection{Function: {\tt largeMPS}, {\tt largeMPO}, {\tt largeLenv}, {\tt largeRenv} and {\tt largeEnv}}

Constructor for the large tensor types ({\it i.e.}, stored on disk) containing MPS, MPO, and environment tensors.

\begin{lstlisting}[language=julia]
function largeMPS(psi::MPS;label::String="mps_",
	names::Array{String,1}=[label*"$i" for i = 1:length(psi)],
	ext::String=file_extension)
  lastnum = 1
  for b = 1:length(psi)
    C = psi[b]
    tensor2disc(names[b],C,ext=ext)
    lastnum *= eltype(C)(1)
  end
  return largematrixproductstate(names,psi.oc,typeof(lastnum))
end

function largeMPS(Ns::Integer;label::String="mps_",
	names::Array{String,1}=[label*"$i" for i = 1:Ns],
	ext::String=file_extension,thisoc::Integer=1,type::DataType=Float64)
  return largematrixproductstate(names,thisoc,type)
end

function largeMPS(type::DataType,Ns::Integer;label::String="mps_",
	names::Array{String,1}=[label*"$i" for i = 1:Ns],
	ext::String=file_extension,thisoc::Integer=1)
  return largeMPS(Ns,label=label,names=names,ext=ext,thisoc=thisoc,type=type)
end

function largeMPO(mpo::P;label::String="mpo_",
	names::Array{String,1}=[label*"$i" for i = 1:length(mpo)],
	ext::String=file_extension) where P <: Union{AbstractArray,MPO}
  lastnum = 1
  for b = 1:length(mpo)
    C = mpo[b]
    tensor2disc(names[b],C,ext=ext)
    lastnum *= eltype(C)(1)
  end
  return largematrixproductoperator(names,typeof(lastnum))
end

function largeMPO(Ns::Integer;label::String="mpo_",
	names::Array{String,1}=[label*"$i" for i = 1:Ns],
	ext::String=file_extension,type::DataType=Float64)
  return largematrixproductoperator(names,type)
end

function largeMPO(type::DataType,Ns::Integer;label::String="mpo_",
	names::Array{String,1}=[label*"$i" for i = 1:Ns],
	ext::String=file_extension)
  return largeMPO(Ns,label=label,names=names,ext=ext,type=type)
end

function largeLenv(lenv::P;label::String="Lenv_",
	names::Array{String,1}=[label*"$i" for i = 1:length(lenv)],
	ext::String=file_extension) where P <: Union{AbstractArray,MPO}
  lastnum = 1
  for b = 1:length(lenv)
    C = lenv[b]
    tensor2disc(names[b],C,ext=ext)
    lastnum *= eltype(C)(1)
  end
  return largeenvironment(names,typeof(lastnum))
end

function largeLenv(Ns::Integer;label::String="Lenv_",
	names::Array{String,1}=[label*"$i" for i = 1:Ns],
	ext::String=file_extension,type::DataType=Float64)
  return largeenvironment(names,type)
end

function largeLenv(type::DataType,Ns::Integer;label::String="Lenv_",
	names::Array{String,1}=[label*"$i" for i = 1:Ns],
	ext::String=file_extension)
  return largeLenv(Ns,label=label,names=names,ext=ext,type=type)
end
export largeLenv

function largeRenv(renv::P;label::String="Renv_",
	names::Array{String,1}=[label*"$i" for i = 1:length(renv)],
	ext::String=file_extension) where P <: Union{AbstractArray,MPO}
  return largeLenv(renv,label=label,names=names,ext=ext)
end

function largeRenv(Ns::Integer;label::String="Renv_",
	names::Array{String,1}=[label*"$i" for i = 1:Ns],
	ext::String=file_extension,type::DataType=Float64)
  return largeenvironment(names,type)
end

function largeRenv(type::DataType,Ns::Integer;label::String="Renv_",
	names::Array{String,1}=[label*"$i" for i = 1:Ns],
	ext::String=file_extension)
  return largeRenv(Ns,label=label,names=names,ext=ext,type=type)
end
export largeRenv

function largeEnv(lenv::P,renv::P;Llabel::String="Lenv_",
	Rlabel::String="Renv_",
	Lnames::Array{String,1}=[Llabel*"$i" for i = 1:Ns],
	Rnames::Array{String,1}=[Rlabel*"$i" for i = 1:Ns],
	ext::String=file_extension,
	type::DataType=Float64) where P <: Union{AbstractArray,MPO}
  return largeLenv(lenv,names=Lnames,type=type),
  			largeRenv(renv,names=Rnames,type=type)
end

function largeEnv(Ns::Integer;Llabel::String="Lenv_",Rlabel::String="Renv_",
	Lnames::Array{String,1}=[Llabel*"$i" for i = 1:Ns],
	Rnames::Array{String,1}=[Rlabel*"$i" for i = 1:Ns],
	ext::String=file_extension,type::DataType=Float64)
  return largeenvironment(Lnames,type),largeenvironment(Rnames,type)
end

function largeEnv(type::DataType,Ns::Integer;Llabel::String="Lenv_",
	Rlabel::String="Renv_",
	Lnames::Array{String,1}=[Llabel*"$i" for i = 1:Ns],
	Rnames::Array{String,1}=[Rlabel*"$i" for i = 1:Ns],
	ext::String=file_extension)
  return largeEnv(Ns,Llabel=Llabel,Lnames=Lnames,Rlabel=Rlabel,Rnames=Rnames,
  			ext=ext,type=type)
end
export largeEnv
\end{lstlisting}

\subsubsection{Elementary functions: {\tt getindex}, {\tt setindex!}, {\tt lastindex}, {\tt length}, and {\tt eltype}}

The standard elementary functions {\tt getindex}, {\tt setindex!}, {\tt lastindex}, {\tt length}, and {\tt eltype} that were used for {\tt tensor.jl} earlier are defined here.  The {\tt type} field is used to avoid reading the tensor from disk in {\tt eltype}.  

\lstinputlisting[language=julia]{code/lib/MPtask/large_elementary.jl}

Note a critical feature of this construction.  When using the {\tt getindex} or {\tt setindex} commands, the tensor is automatically read from the disk.  This is a major time saver when writing algorithms.  The same syntax ({\it i.e.}, {\tt psi[1]}) can be used, avoiding the need to recode each function.  These functions can be used with any tensor type in the library.

\subsection{Function: Load types {\tt loadMPS}, {\tt loadMPO}, {\tt loadLenv}, and {\tt loadRenv}}

Each of these functions loads a given MPS, MPO, or environment from the disk.  All tensors must be saved on the hard disk in the correct filepath (determined by the {\tt label} and {\tt names} fields) before using these functions.  All tensors are assumed to be of a uniform type and that the file extensions are the same between each object as a default.  The only required input to these functions is the number of sites.

\lstinputlisting[language=julia]{code/lib/MPtask/loadlargeTens.jl}

\subsection{Function: {\tt copy} (large types)}

Copying objects for the large tensor types requires that the tensor be read from disk and then given a new filename to place the new object.  Hence, each of these files take a new file name as an argument.

This copy operation for large types is very different from the same operation for dense types. This means that an algorithm that requires this operation will need to take this into account if designed for large types. However, most of the algorithms work equally well with this interface.

\begin{lstlisting}[language=julia]
function copy(names::Array{String,1},X::largeMPS;ext::String=file_extension,
		copyext::String=ext)
  newObj = copy(X)
  newObj.A = names
  for i = 1:length(X)
    Y = tensorfromdisc(names[i],ext=ext)
    tensor2disc(X.A[i],Y,ext=copyext)
  end
  return newObj
end

function copy(names::Array{String,1},X::largeMPO;ext::String=file_extension,
		copyext::String=ext)
  newObj = copy(X)
  newObj.H = names
  for i = 1:length(X)
    Y = tensorfromdisc(names[i],ext=ext)
    tensor2disc(X.H[i],Y,ext=copyext)
  end
  return newObj
end

function copy(names::Array{String,1},X::largeEnv;ext::String=file_extension,
		copyext::String=ext)
  newObj = copy(X)
  newObj.V = names
  for i = 1:length(X)
    Y = tensorfromdisc(names[i],ext=ext)
    tensor2disc(X.V[i],Y,ext=copyext)
  end
  return newObj
end
\end{lstlisting}




%
%

\subsection{MPS gauges: movement of the orthogonality center}

As explained in Ref.~\onlinecite{dmrjulia1}, the MPS can be re-gauged so that the center of orthogonality is contained on another site.  These functions will perform the necessary decompositions to change the gauge of the MPS.

\subsubsection{Function: {\tt moveR} \& {\tt moveR!}}

This function will perform a decomposition on the rank-3 tensor of the MPS and move the orthogonality center one site to the right.  Note that there are two applicable decompositions, {\tt qr} and {\tt svd}.  The {\tt svd} admits truncation so it is used in general.  However, the movement of the orthogonality center often does not need to be truncated as is the case in the bulk of the MPS.  So, when the size of the newly generated index is guaranteed not to be truncated, the {\tt qr} decomposition is used.  When using the {\tt svd}, an extra contraction of $D$ into the new tensor must be performed.  Conversely, the {\tt qr} decomposition requires no extra contraction.

\lstinputlisting[language=julia]{code/lib/MPtask/moveR.jl}

There is also the possibility to update the tensor in-place on the MPS's elements with the {\tt moveR!} function.

\lstinputlisting[language=julia]{code/lib/MPtask/moveR_inp.jl}

\subsubsection{Function: {\tt moveL} \& {\tt moveL!}}

The same rationale for the right move will equally apply to the left movement. However, instead of the {\tt qr} decomposition for the non-truncating operation, the {\tt lq} decomposition is used.

\lstinputlisting[language=julia]{code/lib/MPtask/moveL.jl}

There is also the possibility to update the tensor in-place on the MPS's elements with the {\tt moveL!} function.

\lstinputlisting[language=julia]{code/lib/MPtask/moveL_inp.jl}

\subsubsection{Function: {\tt movecenter!}}

The combination of {\tt moveR} and {\tt moveL} can be combined into a single function {\tt movecenter!} that when given a new orthogonality center will move the center until the new orthogonality center is obtained.

\lstinputlisting[language=julia]{code/lib/MPtask/movecenter_inp.jl}

This function is defined since the main interface function that the user sees can be named more simply.  However, that function can be amended with another core function (like {\tt movecenter!}) if another type of wavefunction ansatz is gauged and must be moved.

\subsubsection{Function: {\tt move!} \& {\tt move}}

The {\tt move!} function is the function that the user will use.  The core operation of moving the orthogonality center can be switched out for another function if necessary.  The tensors are updated in-place.

\lstinputlisting[language=julia]{code/lib/MPtask/move_inp.jl}

The {\tt move} function simply copies the input MPS and returns a completely new MPS.

\subsection{Function: {\tt leftnormalize}, {\tt leftnormalize!}, {\tt rightnormalize}, \& {\tt rightnormalize!}}

The {\tt leftnormalize} and {\tt rightnormalize} function generate the left- or right- normalization of the MPS \cite{dmrjulia1}.  That is, all the tensors are left as the $\hat U$ or $\hat V^\dagger$ tensors simply.  The other elements returned are the $\hat D$ matrix and the last $\hat V^\dagger$ tensors for the case of the left-normalization. The right-normalization returns also the $\hat U$ and $\hat D$ tensors.  The orthogonality center is set to zero after performing this operation.

\lstinputlisting[language=julia]{code/lib/MPtask/leftnormalize.jl}

\subsubsection{Function: {\tt boundarymove!} \& {\tt boundarymove}}

Sometimes it is useful to move the MPS's orthogonality center and then simultaneously update the environment tensors at the same time.  This function {\tt boundarymove!} will do so and manipulate tensors in place.

\lstinputlisting[language=julia]{code/lib/MPtask/boundarymove_inp.jl}

A second function {\tt boundarymove} will copy the input tensor and output a copy of the tensor, leaving the original unchanged.

\lstinputlisting[language=julia]{code/lib/MPtask/boundarymove.jl}

\subsection{Updating and making environment tensors}

DMRjulia uses a specific form for the environment tensors.  The indices are defined for the left and right environments according to the diagram in Ref.~\onlinecite{dmrjulia1}.   This will play a crucial role in several algorithms from here on out.

\subsubsection{Function: {\tt Lupdate} \& {\tt Lupdate!}}

The function {\tt Lupdate} will update the environment tensors in the environment according to the convention in the figure above.

\lstinputlisting[language=julia]{code/lib/MPtask/Lupdate_inp.jl}

The input for {\tt Lupdate!} can be defined with an explicitly defined {\tt dualpsi} ($\langle\psi|$, the dual of the wavefunction) or the wavefunction only ($|\psi\rangle$).  If only the wavefunction is defined, then it is used as the dual wavefunction.  Note that these functions admit any number of MPOs.

\subsubsection{Function: {\tt Rupdate} \& {\tt Rupdate!}}

The function {\tt Rupdate} will update the environment tensors in the environment according to the convention in the figure above.

\lstinputlisting[language=julia]{code/lib/MPtask/Rupdate_inp.jl}

The conventions are listed above and the functions perform similarly to the {\tt Lupdate} functions.

\subsubsection{Function: {\tt makeBoundary}}

This will make the leftmost or right most edge tensor in the network for the environment tensors.

\lstinputlisting[language=julia]{code/lib/MPtask/makeBoundary.jl}

\subsubsection{Function: {\tt makeEnds}}

This function generates the edge tensors in the tensor network for the matrix product state and however many matrix product operators are in use.

\lstinputlisting[language=julia]{code/lib/MPtask/makeEnds.jl}

%
%

%

%
%

\subsubsection{Function: {\tt makeEnv}}

Using the {\tt makeEnds} function to generate the edge tensors, this function wil also perform all contractions between those tensors and the current orthogonality center of the input MPS. This will generate the environment tensors for the system.

\lstinputlisting[language=julia]{code/lib/MPtask/makeEnv.jl}

\subsection{Measurements and related functions}

Some of the most compact and yet most versatile functions will be defined here. The {\tt expect} function will evaluate any correlation function provided the MPS and any number of MPOs.  Similarly, the correlation function evaluates any $N$-point function that is input.  There are efficient algorithms for performing these operations and are described in the following.

\subsubsection{Function: {\tt applyMPO}}

The application of the MPO to the MPS can be accomplished locally.    The diagram in Ref.~\onlinecite{dmrjulia1} conveys the algorithm used. The MPO tensor is contracted onto the MPS. Then, the resulting tensor is decomposed according to the SVD and the result is contracted onto the next tensor.

\lstinputlisting[language=julia]{code/lib/MPtask/applyMPO.jl}

One can use this to contract the MPO onto the MPS, making a resulting that that is $|\mathcal{H}\psi\rangle$.  However, this is ultimately inefficient for making measurements. By using the locality of the MPS, the cost can be brought down considerably.

\subsubsection{Function: {\tt expect}}

The expect function evaluates any correlation of the form $\langle\psi|\mathcal{H}^p|\psi\rangle$ where $p\geq0$ and is an integer, the number of MPOs input. Using a self-similar pattern that emerges in the tensors, the function can be programmed to be very short and concise.  The particular pattern is displayed in Ref.~\onlinecite{dmrjulia1}.

Tracing over the tensors using this pattern results in the best scaling computation to produce the end result.

\lstinputlisting[language=julia]{code/lib/MPtask/expect.jl}

The additional function with only one input MPS is for ease of programming.

\subsubsection{Function: {\tt correlationmatrix}}

For the special case of two-point correlation functions of the form, for example, $\langle\psi|\hat c^\dagger_i\hat c_j|\psi\rangle$, the locality of the orthogonality center can be used to make efficient operations to make the resulting quantities.

\lstinputlisting[language=julia]{code/lib/MPtask/correlationmatrix.jl}

A key point is to use the orthogonality of the MPS to recycle the left or right environment to keep the computational cost down.

\subsubsection{Function: {\tt operator\_in\_order!}}

This function aids the computation of the full correlation function calculator. This function modifies a vector of any size with input sizes representing how many lattice points receive a particular operator. The return value is a vector incremented in the last position. This allows for the evaluation of all operators in order on the lattice. For example, a four-operator correlation function on the indices $i$, $j$, $k$, and $\ell$ sites such that $i\leq j\leq k\leq \ell$ and that the environments are recycled at each step.

\lstinputlisting[language=julia]{code/lib/MPtask/operator_in_order.jl}

\subsubsection{Function: {\tt permutations}}

This is the standard heap algorithm for obtaining permutations of a vector.  The original input in the main {\tt correlation} function is {\tt [i for i = 1:nrank]} for a rank {\tt nrank} tensor \cite{press1992numerical}.

\lstinputlisting[language=julia]{code/lib/MPtask/permutations.jl}

\subsubsection{Function: {\tt correlation}}

This function evaluates any correlation function with any number of operators.  For example, $\langle\hat c^\dagger_i\hat c^\dagger_j\hat c_k\hat c_\ell\rangle$. The operation is kept efficient by recycling the environment tensors for a given order of indices, for example $i\leq j\leq k\leq \ell$. Then, the {\tt permutations} function is used to change the order and the computation is repeated.  There is no assumed symmetry in the resulting tensor output. For a symmetric evaluation of a two-operator expression, see {\tt correlationmatrix}.

\lstinputlisting[language=julia]{code/lib/MPtask/correlation.jl}

\subsection{Constructing MPOs and MPSs}

Constructing the MPOs and MPSs can be tedious using basic operations.  These functions make it easier to define the MPO. The convention used for MPSs is given in Ref.~\onlinecite{dmrjulia1} as is the convention for the MPOs.


In a subsequent work, the automatic generation the MPO will be used to make these quantities.

\subsubsection{Function: {\tt makeMPO}}

This function converts the MPO form in a systematic way.  The input MPO tensors can be entered as a function or as a vector of rank-2 tensors. If rank-4 tensors are made, then they can be loaded into the MPO easily by defining {\tt MPO(rank4s)}. See Ref.~\onlinecite{dmrjulia1} for more details on writing MPO functions. There are two inputs that could be accommodated for the {\tt makeMPO} function.  One is a simple matrix of numbers.  Operators can be found from the {\tt Opt.jl} module.

Alternatively, the input matrix could be encoded in a function whose only argument is the site index.  

In order to properly convert the rank-2 flattened MPO format, the physical index size must be known, hence the {\tt physindsize} parameter which can be constant for each site (integer input) or vary with a vector of inputs.  The vector will repeat on, for example, an input of 3-site vector for a 10-site lattice will assume that a modular assignment is taken on the lattice.

\lstinputlisting[language=julia]{code/lib/MPtask/makeMPO.jl}

The default function assumes that the MPO was written in the lower triangular form.  If the upper triangular form is used, then the flag {\tt lower} must be set to {\tt false}.  The regular julia Arrays can be used instead of the {\tt denstens} type by setting {\tt regtens=true}.

\subsubsection{Function: {\tt makeMPS}}

Given a vector that could result from an exact diagonalization computation, the {\tt makeMPS} function converts to the MPS form.  To find the MPS form, the size of the physical index must be input, {\tt physInd}, in addition to the vector.

\lstinputlisting[language=julia]{code/lib/MPtask/makeMPS.jl}

Note that the {\tt qr} (or {\tt lq}) decomposition is used, expanding on the method in Ref.~\onlinecite{bakerCJP21}.

\subsubsection{Function: {\tt penalty!} \& {\tt penalty}}

In order to find excitations on the MPS, there is a strategy to add a penalty term to the MPO based on the ground state of the form
\begin{equation}
\mathcal{H}=\mathcal{H}_0+\lambda|\psi_0\rangle\langle\psi_0|
\end{equation}
where $\psi_0$ is the ground state found from $\mathcal{H}_0$. Given the penalty $\lambda$, the new "ground state" will be the first excitation. This method can suffer from poor convergence issues and vastly increasing bond dimension of the MPO.  There are many other ways to compute the excitations that produce higher quality results and are more efficient.


\lstinputlisting[language=julia]{code/lib/MPtask/penalty.jl}

\subsubsection{Function: {\tt transfermatrix}}

One very powerful, but expensive and not commonly used, analysis tools in a tensor network is the determination of the correlation length from the transfer matrix.  Ref.~\onlinecite{dmrjulia1} contains more information on this technique.  The function contracts the link indices between the input indices $i$ and $j$ to obtain the transfer matrix.  The function can be reused by placing the output transfer matrix back into the function with the optional {\tt transfermat} option to the function.

\lstinputlisting[language=julia]{code/lib/MPtask/transfermatrix.jl}

\subsection{Converting to and from exact diagonalization computations}

\subsubsection{Function: {\tt fullpsi}}

The function {\tt fullpsi} contracts all link indices in the MPS and returns the full wavefunction suitable for exact diagonalization.

\lstinputlisting[language=julia]{code/lib/MPtask/fullpsi.jl}

\subsubsection{Function: {\tt fullH}}

The function {\tt fullH} will contract the entire MPO and produce a matrix representing the Hamiltonian. The function simply contracts the link indices and reorders the remaining physical indices so that the matrix is properly represented.

\lstinputlisting[language=julia]{code/lib/MPtask/fullH.jl}

This method will only work for a few sites before running into memory problems on the classical computer.  The exact number depends on the size of the physical index of each site.  Using double precision allows for approximately 10 Lanczos coefficients to be determined accurately. Upgrading the computation to handle quadruple precision is possible but requires a (slower) eigenvalue solver capable of handling this.

\end{widetext}

\end{document}